\documentclass[pdflatex,sn-nature]{sn-jnl}

\usepackage{graphicx}%
\usepackage{multirow}%
\usepackage{amsmath,amssymb,amsfonts}%
\usepackage{amsthm}%
\usepackage[title]{appendix}%
\usepackage{xcolor}%
\usepackage{textcomp}%
\usepackage{manyfoot}%
\usepackage{booktabs}%
\usepackage{hyperref}
\usepackage[normalem]{ulem}
\let\orcidlogo\relax
\usepackage{orcidlink}
\usepackage{longtable}
\usepackage{algorithm}%
\usepackage{algorithmicx}%
\usepackage{algpseudocode}%
\usepackage{listings}%
\usepackage{adjustbox}
\usepackage{geometry}
\usepackage{etoolbox}

\newcommand{\sobs}{\ifmmode {\sigma_{\parallel,{\rm obs}}} \else  {$\sigma_{\parallel,{\rm obs}}$} \fi}
\newcommand{\slen}{\ifmmode {\sigma_{\parallel,{\rm len}}} \else  {$\sigma_{\parallel,{\rm len}}$} \fi}
\newcommand{\sobsj}{\ifmmode {\sigma_{\parallel,{\rm obs}}^j} \else  {$\sigma_{\parallel,{\rm obs}}^j$} \fi}
\newcommand\farcs{\mbox{$.\!\!^{\prime\prime}$}}%
\newcommand{\mc}[1]{\multicolumn{1}{c}{#1}}

\begin{document}

\title{Joint constraints on gravity and stellar orbital anisotropy in massive galaxies}


\author*[1]{\fnm{Wei} \sur{Du}\,\orcidlink{0000-0001-9781-6863}}\email{duwei@shnu.edu.cn}

\author*[1]{\fnm{Liping} \sur{Fu}\,\orcidlink{0000-0003-0688-8445}}\email{fuliping@shnu.edu.cn}

\author*[2]{\fnm{Gong-Bo} \sur{Zhao}\,\orcidlink{0000-0003-4726-6714}}\email{gbzhao@nao.cas.cn}

\author[3]{\fnm{Yiping} \sur{Shu}\,\orcidlink{0000-0002-9063-698X}}
\author[2]{\fnm{Shuo} \sur{Yuan}}
\author[4]{\fnm{Zuhui} \sur{Fan}\,\orcidlink{0000-0002-8397-012X}}
\author[5]{\fnm{Huanyuan} \sur{Shan}\,\orcidlink{0000-0001-8534-837X}}
\author[1]{\fnm{Chenggang} \sur{Shu}\,\orcidlink{0000-0003-1055-2039}}

\affil*[1]{\orgdiv{Shanghai Key Lab for Astrophysics}, \orgname{Shanghai Normal University}, \orgaddress{\city{Shanghai}, \postcode{200234}, \country{China}}}

\affil[2]{\orgdiv{National Astronomical Observatories}, \orgname{Chinese Academy of Science}, \orgaddress{\city{Beijing}, \postcode{100101}, \country{China}}}

\affil[3]{\orgdiv{Purple Mountain Observatory}, \orgname{Chinese Academy of Science}, \orgaddress{\city{Nanjing}, \postcode{210023}, \country{China}}}

\affil[4]{\orgdiv{South-Western Institute for Astronomy Research}, \orgname{Yunnan University}, \orgaddress{\city{Kunming}, \postcode{650500}, \country{China}}}

\affil[5]{\orgdiv{Shanghai Astronomical Observatory}, \orgname{Chinese Academy of Science}, \orgaddress{\city{Shanghai}, \postcode{200030}, \country{China}}}

\abstract{Strong gravitational lensing combined with stellar dynamics provides a complementary route for testing gravity on kiloparsec scales and probing the internal structure of massive galaxies. However, such studies remain limited by degeneracies among the mass-density profile, stellar orbital anisotropy and external convergence, and by modelling assumptions, especially when only single-aperture velocity dispersions are available. Here we develop a hierarchical Bayesian framework to disentangle gravity and stellar orbital anisotropy from other effects at the population level. By reconstructing the lens mass distribution with a flexible broken power-law model and propagating its posterior uncertainty into the predicted velocity dispersion, we obtain a likelihood for each lens in the plane of stellar orbital anisotropy and an effective mismatch parameter. This parameter encapsulates projection bias, external convergence, cosmological distance ratios and deviations from general relativity via the post-Newtonian parameter $\gamma_{\rm PPN}$. Applying this framework to 121 galaxy-scale lenses, we find $\gamma_{\rm PPN}=1.027^{+0.099}_{-0.095}$, consistent with general relativity, and obtain $2\sigma$ evidence that the stellar orbits of massive galaxies have become more radially biased over the past $\sim6$ Gyr. Forecasts show that future samples of order $10^5$ lenses could enable sub-percent tests of gravity, precise measurements of orbital-structure evolution and complementary constraints on the cosmological matter-density parameter.}


\maketitle

\section{Introduction}\label{sec1}
The combination of strong gravitational lensing and stellar dynamics provides a powerful way to test gravity on kiloparsec scales and to probe the internal structure of massive galaxies \cite{2006PhRvD..74f1501B,2009ApJ...703L..51K,2010ApJ...708..750S,2018Sci...360.1342C,2023MNRAS.521.6005E}. In practice, however, its constraining power is limited by two major degeneracies: the mass-sheet degeneracy (MSD) in lens mass modelling \cite{1985ApJ...289L...1F,2013A&A...559A..37S} and the mass-anisotropy degeneracy (MAD) in dynamical modelling \cite{1982MNRAS.200..361B,2026ApJ..1000..264V}. The MSD arises because adding a uniform mass sheet and rescaling the lens mass distribution can leave the lensed image configuration unchanged, while rescaling the source plane and the predicted time delays. For an individual lens, the MSD can be partly restricted by adopting a specific mass-density profile and accounting for an effective external convergence at the lens redshift \cite{2020A&A...643A.165B}. The latter is usually estimated indirectly using weak gravitational lensing \cite{2018MNRAS.477.5657T} or weighted galaxy counts along the line of sight \cite{2017MNRAS.467.4220R,2020MNRAS.498.1420W}. The MAD arises because different combinations of the mass density profile and stellar orbital anisotropy can yield very similar line-of-sight stellar velocity dispersions. Spatially resolved kinematics from integral-field spectroscopy can help alleviate this degeneracy, but such data are currently available only for a small number of lenses, mostly at low redshift \cite{2024ApJ...970...86S,2026arXiv260412155K}. For the majority of known strong lensing systems, the dynamical information remains limited to single-aperture velocity-dispersion measurements, which by themselves provide little constraint on stellar orbital anisotropy. This is the observational regime considered in this work.

Consequently, joint strong-lensing and stellar-dynamical (L\&D) analyses cannot fully separate the effects of the MSD and MAD in an individual lens system based solely on lensed images and single-aperture stellar kinematics. Previous studies have therefore relied on informative priors or simplifying assumptions about the main lens mass distribution, the external convergence $\kappa_{\rm ext}$, and the stellar orbital anisotropy $\beta$. For instance, the single power-law model has been widely adopted as a baseline model for the mass-density profile of galaxy-scale lenses \cite{2026arXiv260412155K}. External convergence has often been ignored in L\&D analyses, equivalent to assuming $\kappa_{\rm ext}=0$, whereas it is routinely considered in time-delay cosmography because it directly affects the inferred time-delay distance \cite{2017MNRAS.468.2590S,2020A&A...639A.101M}. Stellar orbital anisotropy has commonly been modelled with simple parametric forms \cite{2026ApJ..1000..264V}, such as a constant $\beta$ with informative priors motivated by local early-type galaxies \cite{2001AJ....121.1936G,2006PhRvD..74f1501B}, or an Osipkov--Merritt profile \cite{1979PAZh....5...77O,1985AJ.....90.1027M}. These choices have enabled specific applications, but they also make the resulting constraints conditional on adopted priors or assumptions. One indication of such model dependence is that lensing-only and L\&D studies have reported opposite trends in the redshift evolution of the total mass-density slopes of lens galaxies \cite{2012ApJ...757...82B,2023MNRAS.521.6005E,2025A&A...694A.196G}, suggesting the presence of residual systematics in lens and dynamical modelling \cite{2017MNRAS.464.3742R,2024ApJ...970...86S}.

Conventionally, L\&D analyses use the Einstein radius inferred from lensed images, together with assumptions about the lens mass distribution, stellar light distribution and stellar orbital anisotropy, to predict an observable single-aperture line-of-sight (LOS) velocity dispersion $\sigma_{\parallel,\mathrm{len}}$ for each lens \cite{2010ApJ...708..750S,2012ApJ...757...82B,2017ApJ...835...92C,2019MNRAS.488.3745C}. Parameters of interest are then inferred by comparing this prediction with the observed LOS velocity dispersion $\sigma_{\parallel,\mathrm{obs}}$ from single fibre spectroscopy. This strategy is well motivated because the Einstein radius is a robust lensing observable, but the comparison itself is sensitive to several coupled ingredients. A mismatch between $\sigma_{\parallel,\mathrm{len}}$ and $\sigma_{\parallel,\mathrm{obs}}$ may arise from the assumed radial mass profile, external convergence, projection bias in the dynamical modelling, deviations from general relativity, differences between the true and adopted cosmological distance ratios, or stellar orbital anisotropy. A robust inference therefore requires these ingredients to be modelled, calibrated or marginalized in a consistent framework.%

To address this requirement, we propagate the joint posterior distribution of the lens mass parameters into the prediction of the single-aperture velocity dispersion, rather than anchoring the dynamical prediction to the Einstein radius. For the lens mass modelling, we adopt a flexible broken power-law (BPL) model validated by hydrodynamical simulations \cite{2020ApJ...892...62D,2023ApJ...953..189D}. This model has been shown to recover the projected mass distribution from the galaxy centre to several Einstein radii with controlled biases, under physically motivated priors on the central region and the outer density slope. For a more self-consistent dynamical model, we directly fit the stellar light distribution with a flexible power-law--S\'ersic profile, matched to the BPL mass model. We treat the external convergence as the sum of contributions from the lens environment and line-of-sight large-scale structures, and quantify the projection bias induced by spherical dynamical modelling. These steps provide a controllable uncertainty budget for the lensing-predicted velocity dispersion of each lens.

After marginalizing over the uncertainty in the lens mass-density profile, each lens yields a likelihood in the $X$--$\beta$ plane. The parameter $X$ quantifies the mismatch between $\sigma_{\parallel,\mathrm{obs}}$ and $\sigma_{\parallel,\mathrm{len}}$, absorbing projection bias, external convergence, deviations from general relativity through the parameterized post-Newtonian parameter $\gamma_{\mathrm{PPN}}$, and the dependence on cosmological distance ratios, while $\beta$ remains an explicit internal dynamical parameter controlling the stellar orbital structure. Although $X$ and $\beta$ remain degenerate for any individual lens, combining many such likelihoods in a hierarchical Bayesian framework allows their population-level behaviour to be inferred. Applying this framework to a sample of 121 galaxy-scale lenses, we constrain $\gamma_{\mathrm{PPN}}$ and the redshift evolution of anisotropy, $\beta(z)=\beta_0+\beta_z z$, together with its intrinsic scatter $\tau_\beta$. Forecasts show that future samples can substantially tighten constraints on $\gamma_{\mathrm{PPN}}$ and the evolution of $\beta$, while also enabling complementary constraints on $\Omega_{\mathrm{m}}$ and informative, albeit weaker, constraints on the population parameters governing external convergence.

\section{Results}\label{sec2}
\subsection*{Single-lens degeneracies in the $X$--$\beta$ plane}

\begin{figure}
    \centering
    \includegraphics[width=\linewidth]{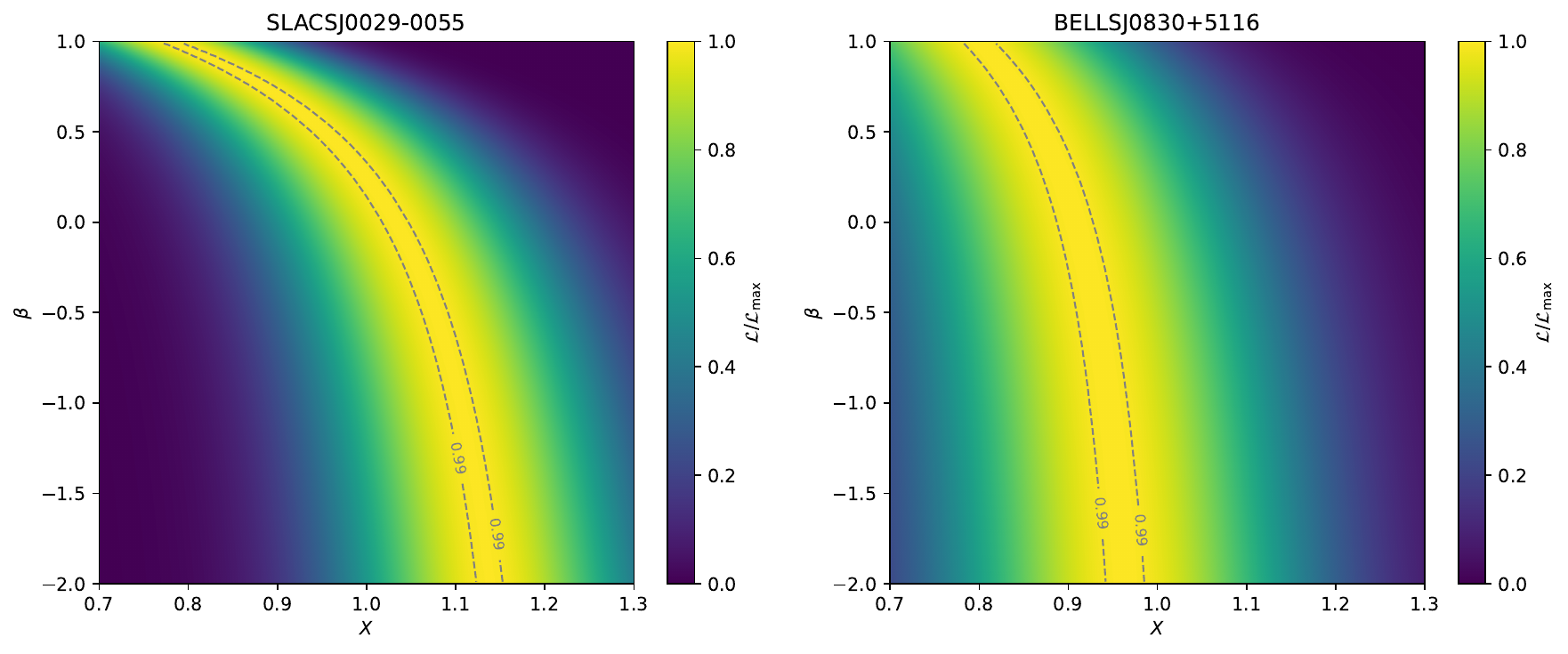}
    \caption{Likelihood maps in the $X$--$\beta$ plane, where $X$ is the effective mismatch parameter and $\beta$ is the stellar orbital anisotropy. Colours show the likelihood $\mathcal{L}(X,\beta)$ normalized by its maximum value for two representative lenses, SLACSJ0029-0055 and BELLSJ0830+5116, after marginalizing over the uncertainty in lens radial mass-density profiles. The elongated colour distributions illustrate the degeneracy between $X$ and $\beta$. Dashed contours mark the $\mathcal{L}/\mathcal{L}_{\max}=0.99$ levels.}
    \label{fig:xbeta}
\end{figure}

We first examine the information provided by L\&D analyses for individual lens systems. For each lens, we obtain a two-dimensional likelihood in the $X$--$\beta$ plane by marginalizing over the joint posterior distribution of the free radial mass-density parameters, the outer slope $\alpha$ and amplitude $b$ (see Fig.~\ref{figA1:bpl_fitting} for examples of lens modelling). Fig.~\ref{fig:xbeta} visualizes the resulting likelihoods for two representative systems. In both cases, the high-likelihood region follows a curved ridge, exhibiting a pronounced degeneracy between $X$ and $\beta$: changes in $X$ can be compensated by changes in $\beta$ while producing comparable single-aperture LOS velocity dispersions. While the strength and shape of the $X$--$\beta$ degeneracy vary from lens to lens, the two parameters consistently show an anti-correlation. This lens-specific degeneracy reflects the difficulty of using any single lens system to uniquely disentangle stellar orbital anisotropy from the physical components encapsulated within $X$. These individual-lens likelihoods therefore serve as the basic data products for the hierarchical Bayesian analysis below, where $X$ and $\beta$ are linked across the lens sample through shared population parameters. 

As detailed in the Methods, the mismatch factor $X$ absorbs the projection bias ($b_\sigma$), the external convergence ($\kappa_{\mathrm{ext}}$), variations in cosmological distance ratios, and any deviations from general relativity parametrized by $\gamma_{\mathrm{PPN}}$. By combining the individual likelihoods $\mathcal{L}(X,\beta)$ from 121 lens systems over the redshift range $0.06$ to $0.66$, the hierarchical Bayesian framework enables a statistical separation of stellar orbital anisotropy from $\gamma_{\rm PPN}$ and other physical effects encoded in $X$ at the population level.

\subsection*{Population constraints on gravity and anisotropy evolution}
\begin{figure}
    \centering
    \includegraphics[width=\linewidth]{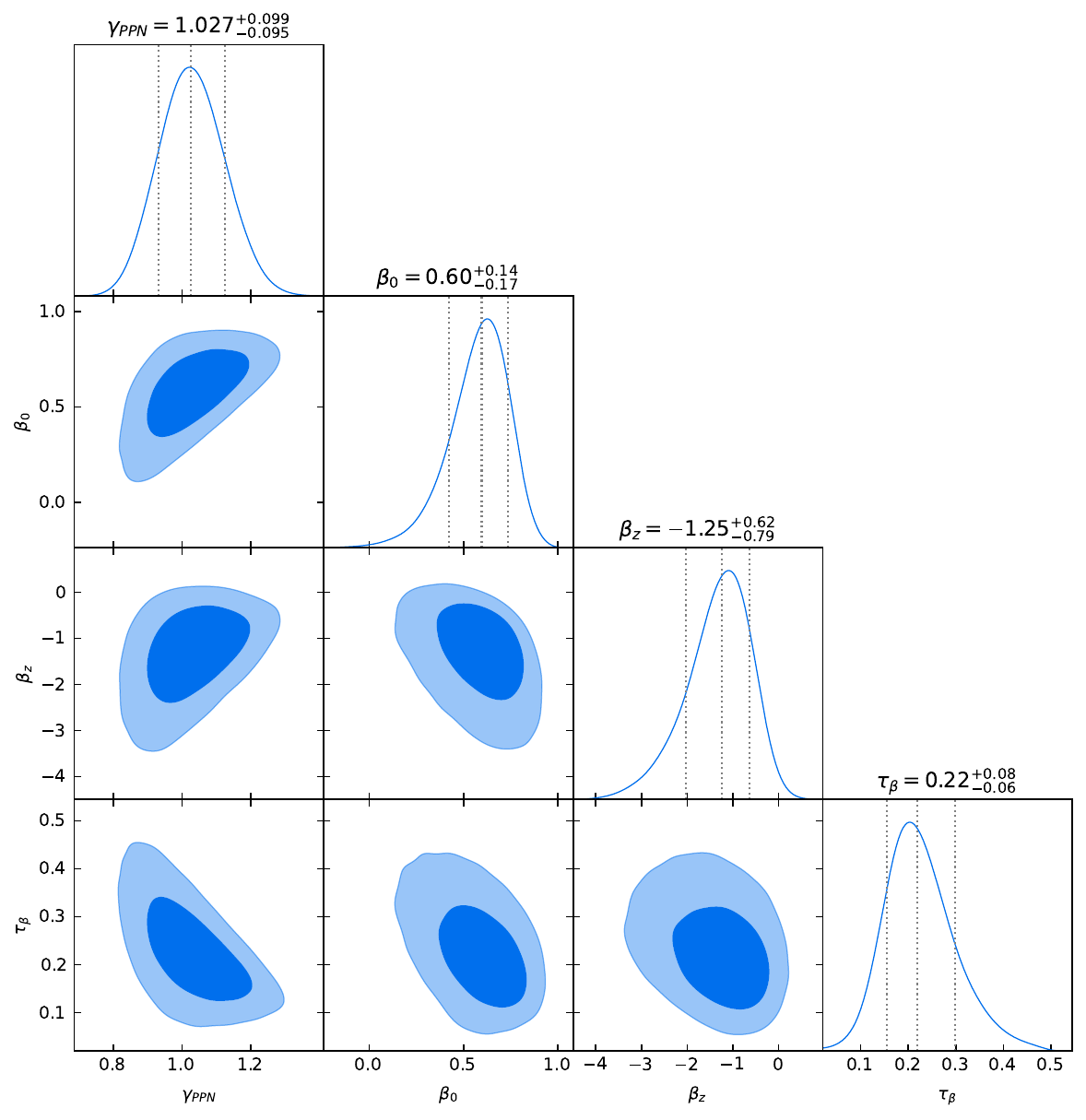}
    \caption{Population-level constraints on the parameterized post-Newtonian parameter and stellar orbital anisotropy from 121 galaxy-scale lenses. Diagonal panels show the one-dimensional marginalized posterior distributions, with the median and 68\% credible interval marked by vertical dotted lines and reported above each panel. Off-diagonal panels show the two-dimensional posteriors, with contours enclosing the 68\% and 95\% credible regions. For clarity, the posterior distributions of $\delta_{m,0}$ and $\delta_{m,z}$ are not shown here; the full corner plot including additional parameters is provided in Fig.~\ref{figA2:mcmc_lam}.}
    \label{fig:mcmc_main}
\end{figure}

Using the fiducial prior on external convergence, corresponding to $\lambda=1$ (see Eq.~\ref{eq:bayes}), we infer the population parameters from the full sample of 121 galaxy-scale lenses. Fig.~\ref{fig:mcmc_main} shows the marginalized posterior distributions of the parameters that are most directly constrained by the current sample. We obtain $\gamma_{\mathrm{PPN}}=1.027\pm0.097$, consistent with the value of $\gamma_{\rm PPN}=1$ for general relativity. We find a positive value of $\beta_0=0.60^{+0.14}_{-0.17}$, indicating radially biased stellar orbits towards the low-redshift end of the population. The slope is negative, $\beta_z=-1.25^{+0.62}_{-0.79}$, favouring a decrease in the mean anisotropy with increasing redshift, or equivalently more radially biased stellar orbits towards later cosmic times. The posterior probability that $\beta_z<0$ is approximately $98.7\%$. We also infer $\tau_\beta=0.22^{+0.08}_{-0.06}$, indicating a non-negligible intrinsic scatter in stellar orbital anisotropy around the mean redshift trend. The gravity constraint and the anisotropy-evolution signal are therefore obtained simultaneously rather than by fixing either component.

\begin{figure}
    \centering
    \includegraphics[width=\linewidth]{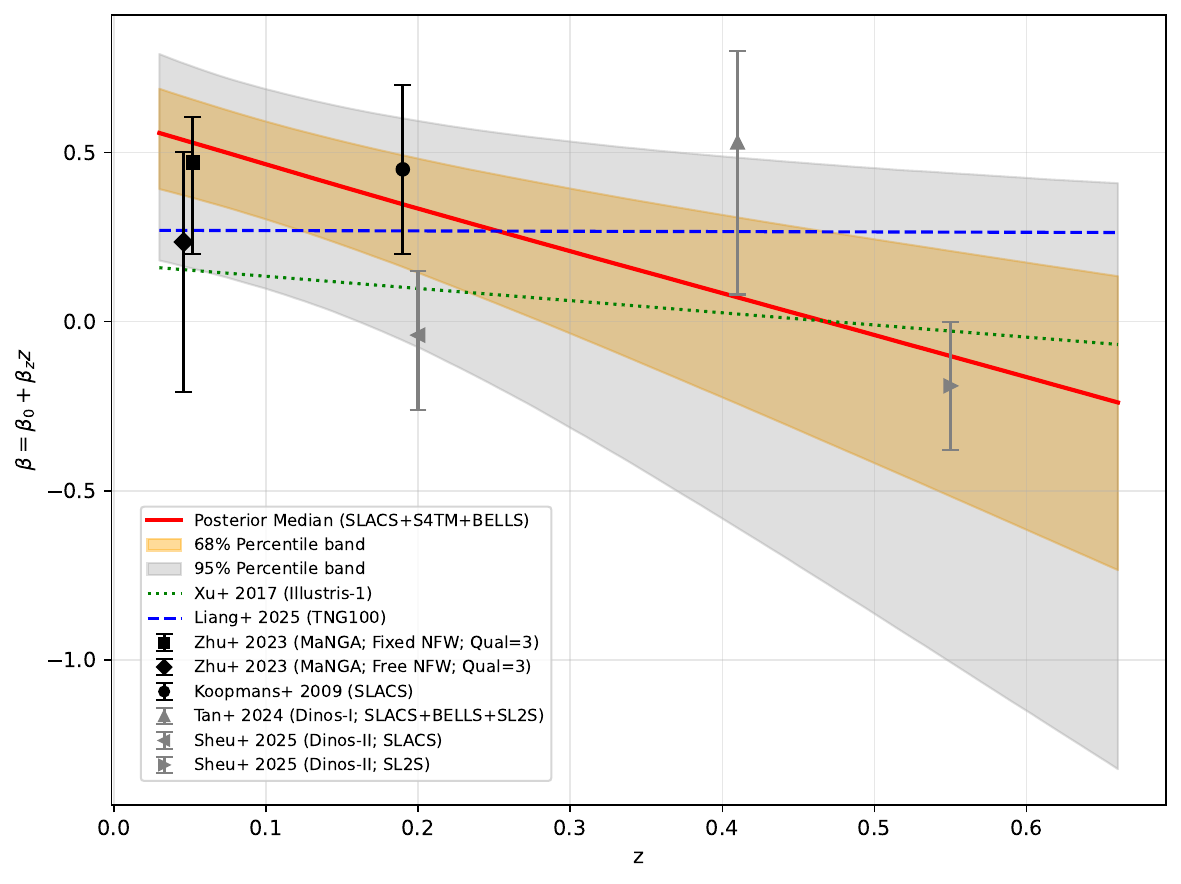}
    \caption{Redshift evolution of stellar orbital anisotropy.
    The red line shows the posterior median of the mean relation $\beta(z)=\beta_0+\beta_z z$ based on the posterior samples of $\beta_0$ and $\beta_z$, with orange and grey shaded regions denoting the 68\% and 95\% posterior percentile bands, respectively. The green dotted and blue dashed lines are the evolutionary trends derived from the Illustris-1 and TNG100 hydrodynamical simulations, respectively. Symbols show literature estimates of $\beta$ for observed massive galaxies, including local galaxies from MaNGA and lens galaxies from SLACS, S4TM, BELLS and SL2S surveys, as labelled. The two MaNGA points correspond to spherical JAM models with fixed and free NFW halo parameters, respectively.}
    \label{fig:beta_z}
\end{figure}

We translate the posterior constraints on $\beta_0$ and $\beta_z$ into the mean redshift trend of stellar orbital anisotropy. Fig.~\ref{fig:beta_z} visualizes the corresponding prediction for the redshift evolution of $\beta$, including the 68\% and 95\% percentile bands. The inferred trend points to more radially biased stellar orbits in massive galaxies at lower redshift. We compare this inferred trend with previous observation- and simulation-based estimates compiled from the literature. This comparison shows that, given the current sample size, our inferred trend is broadly compatible with the literature estimates compiled here, while also illustrating the sensitivity of anisotropy estimates to data sets and modelling assumptions. Examples include the difference between the two MaNGA estimates obtained under fixed and free NFW halo assumptions, the difference between the Dinos-I and Dinos-II population constraints, and the difference between the Illustris-1 and TNG100 trends reflecting their distinct feedback prescriptions and sub-grid physics. Our framework combines the individual $X$--$\beta$ likelihoods of 121 lenses into a hierarchical model, allowing $\gamma_{\mathrm{PPN}}$ and the redshift evolution of $\beta$ to be constrained simultaneously at the population level. The comparison should be read as contextual rather than as a homogeneous meta-analysis, because the definitions and apertures of $\beta$ differ across the studies.

\subsection*{Sensitivity and validation tests}
We further perform a series of sensitivity and validation tests to assess the robustness of our population-level inference. The sensitivity tests, summarized in Section~\ref{secA2:sensitivity_test}, examine the dependence of the inferred parameters on the strength of the external-convergence prior, sample selection and assumed background cosmology. As shown in Table~\ref{tabA1:sensitivity}, the resulting posterior estimates of the post-Newtonian and velocity anisotropy parameters are consistent across the considered $\lambda$ values, sub-samples, and cosmological choices, supporting the robustness of our conclusions. In Section~\ref{secA3:validation_test}, we validate the hierarchical Bayesian inference framework using mock dynamical data generated for the observed lens sample. These tests confirm that the input gravity and anisotropy parameters can be reliably recovered within the statistical uncertainties, although realization-to-realization variations are still evident for the current sample size.

\subsection*{Forecasts for future lens samples}
\begin{figure}[!ht]
    \centering
    \includegraphics[width=\linewidth]{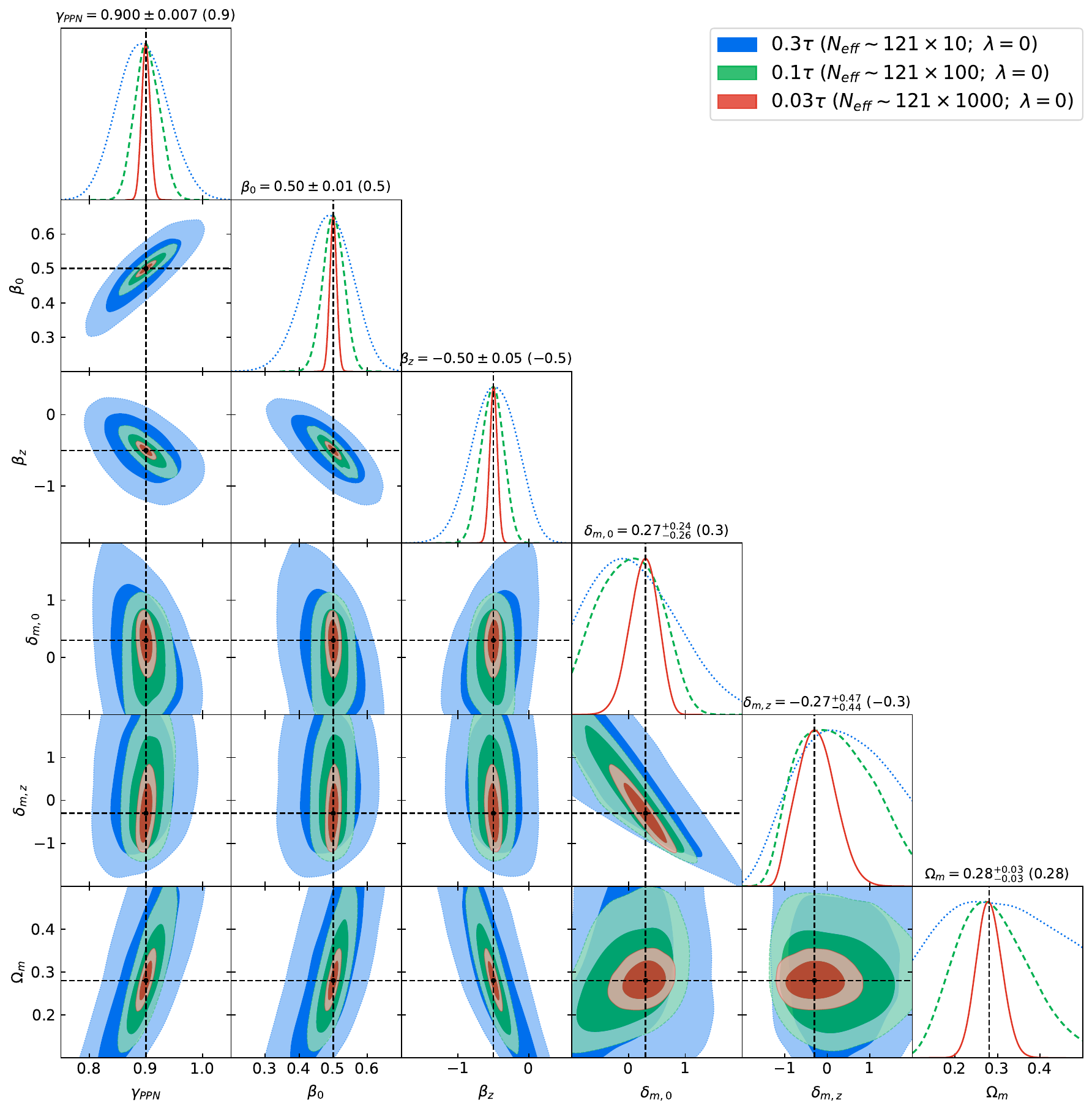}
    \caption{Forecasts for future lens samples based on noise-free mock velocity-dispersion data. The mock data are generated with $\gamma_{\rm PPN}=0.9$, $\beta(z)=0.5-0.5z$, $\delta_m(z)=0.3-0.3z$ and $\Omega_{\rm m}=0.28$ for a flat universe, as indicated by the black dashed lines and by the values in parentheses above the diagonal panels. Diagonal panels show the one-dimensional marginalized posteriors, with the median and 68\% credible interval reported above each panel for the red curves. Off-diagonal panels show the two-dimensional marginalized posteriors, with contours enclosing the 68\% and 95\% credible regions. The blue, green and red curves and contours show the constraints for $0.3\tau$, $0.1\tau$ and $0.03\tau$, respectively, corresponding to effective sample sizes of $N_{\rm eff}\sim121\times10$, $121\times100$ and $121\times1000$. Here, $\tau$ is the total velocity-dispersion uncertainty for each lens, which is rescaled to mimic the statistical gain from larger samples.}
    \label{fig:mcmc_won_b}
\end{figure}

With 121 lenses, the present analysis constrains $\gamma_{\rm PPN}$ at the 10\% level and provides a $2\sigma$ indication of redshift evolution in $\beta$. However, the constraints on the other population parameters remain very weak. Validation tests also show that current-size lens samples provide little constraint on the cosmological matter density parameter $\Omega_{\rm m}$, especially when no tight prior on external convergence is imposed (see Fig.~\ref{figA5:mcmc_won}). This situation will improve substantially with ongoing and next-generation surveys. Over the next 5--10 years, surveys such as Euclid, the Vera C. Rubin Observatory LSST, and CSST are expected to discover $\mathcal{O}(10^5)$ galaxy-scale strong lenses \cite{2015ApJ...811...20C,2024MNRAS.533.1960C}. Such samples will greatly increase the statistical power of lensing--dynamical analyses. 

We therefore use the noise-free mock dynamical data to explore the performance of our hierarchical Bayesian framework for larger lens samples. The mock dynamical data are generated as described in Section~\ref{secA3:validation_test} and are combined with the input lens mass distributions to construct the single-lens $X$--$\beta$ likelihoods. The resulting likelihoods are then analysed using the same hierarchical Bayesian framework as for the observed sample. The statistical gain expected from larger samples is mimicked by rescaling the total velocity-dispersion uncertainty, $\tau$ (Eq.~\ref{eq:psigma}), by a factor of $\sqrt{121/N_{\rm eff}}$, where $N_{\rm eff}$ denotes the effective number of lenses.

Fig.~\ref{fig:mcmc_won_b} shows that the constraints on $\gamma_{\rm PPN}$ and the anisotropy parameters tighten rapidly as the effective sample size increases. For the most optimistic case shown here, corresponding to $\sim121,000$ lenses, the input values of $\gamma_{\rm PPN}$, $\beta_0$ and $\beta_z$ are accurately recovered with tight credible intervals. The precision on $\gamma_{\rm PPN}$ reaches the sub-percent level. The matter-density parameter is also constrained to an accuracy of $\sim0.03$, demonstrating that large samples of galaxy-scale lenses can provide complementary cosmological information from lensing--dynamical analyses.

The external-convergence population parameters remain less tightly constrained than the other parameters, reflecting the limited sensitivity of lensing--dynamical constraints to the population-level properties of line-of-sight large-scale structure. Nevertheless, the constraints on $\delta_{m,0}$ and $\delta_{m,z}$ improve substantially with sample size and begin to recover the input density-contrast relation. These forecasts suggest that future large lens samples will not only sharpen tests of gravity, improve measurements of stellar orbital anisotropy evolution and tighten constraints on $\Omega_{\rm m}$, but also enable informative constraints on the redshift dependence of the line-of-sight density contrast associated with external convergence. However, These forecasts capture only the statistical gains from increasing the effective sample size while holding the selection and properties of the current lens sample fixed. They therefore do not account for the additional constraining power potentially offered by a broader lens-redshift range or for practical limitations in achieving sufficiently precise lens mass modelling and stellar velocity-dispersion measurements. 

\section{Discussion}\label{sec:discussion}
We have developed a hierarchical Bayesian framework to alleviate degeneracies in joint lensing and dynamical analyses. In this framework, the residual degeneracy in each lens is represented by a likelihood in the $X$--$\beta$ plane. The parameter $X$ absorbs projection bias, external convergence, cosmological distance ratios and possible deviations from general relativity, while the anisotropy $\beta$  is described by a population relation $\beta(z)=\beta_0+\beta_z z$ with intrinsic scatter $\tau_\beta$. Thus, the degeneracy that is difficult to resolve for individual lenses is converted into a population-level inference problem. 

The inferred value of $\gamma_{\rm PPN}$ is consistent with general relativity at the 10\% level, thereby providing a galaxy-scale test of gravity. If general relativity holds on these scales, this agreement also indicates that no large unmodelled multiplicative bias is needed to reconcile the lensing-predicted and observed velocity dispersions. Future samples could move beyond this population-averaged test by searching for trends of $\gamma_{\rm PPN}$ with lens properties. In screened modified-gravity models, departures from general relativity may depend on mass and environment \cite{2011PhRvL.107g1303Z}. Large lens samples with well-characterized environments could therefore test whether $\gamma_{\rm PPN}$ varies with lens mass or local environment, and help distinguish between different screening mechanisms on galaxy scales.

The inferred negative value of $\beta_z$ indicates that the mean stellar orbital anisotropy of massive lens galaxies becomes more radially biased towards lower redshift. This trend is qualitatively consistent with the two-phase scenario of massive-galaxy formation \cite{2010ApJ...725.2312O,2023MNRAS.525.2789D}, in which early dissipative growth is followed by later accretion of ex situ stars through gas-poor mergers. The later processes can increase the fraction of stars on radial orbits and modify the global dynamical structure of massive galaxies. A related line of evidence comes from measurements of the kinematic kurtosis $h_4$, which is the fourth Gauss--Hermite moment of the line-of-sight stellar velocity distribution and has been linked to ex situ stellar fractions and orbital anisotropy \cite{1993MNRAS.265..213G,2022MNRAS.513.6134W}. Ref.~\cite{2023MNRAS.525.2789D} reported an increase in $h_4$ towards lower redshift, suggesting an evolving $\beta$ broadly consistent with our direct constraint on $\beta(z)$. A quantitative physical interpretation of $\beta(z)$ will require high-resolution hydrodynamical simulations to calibrate its dependence on merger history and baryonic feedback. Such calibration would allow observational measurements of $\beta(z)$ to be used as a dynamical probe of the assembly history and feedback processes of massive galaxies. 

We have examined whether the inferred anisotropy evolution could be driven by redshift-dependent mass selection. Because more massive galaxies tend to have more radially biased stellar orbits \cite{2017MNRAS.469.1824X}, a preference for more massive lenses at lower redshift could mimic part of the observed trend. In our sample, however, the correlation coefficient between lens redshift and velocity dispersion, used here as a rough proxy for galaxy mass, is only 0.09. This suggests that the inferred evolution of $\beta$ is unlikely to be dominated by a simple mass-related selection effect. A full treatment of the joint selection in redshift, velocity dispersion and lensing cross-section should nevertheless be revisited for future larger samples.

The present analysis relies on spherical Jeans modelling with a constant anisotropy parameter, which is interpreted as the stellar-mass-weighted anisotropy. For the BPL-based modelling adopted here, ref.~\cite{2020ApJ...892...62D} identified a projection-related bias $b_\sigma$ using Illustris-1 galaxies placed at $z\sim0.18$, close to the median redshift of the SLACS lenses. This projection bias is adopted as the fiducial modelling bias in this work. However, a more general modelling bias may additionally depend on the physical size of the spectroscopic aperture and the intrinsic radial profile of the orbital anisotropy. The modelling bias may therefore be redshift dependent and potentially propagate into the inferred evolution of $\beta$. A test using Illustris-1 galaxies at $z=0$ indicates that the modelling bias factor for massive galaxies is typically greater than unity and tends to decrease as the physical aperture size increases. If this trend also applies to the observed lenses, the variation in physical aperture size with redshift would induce a redshift-dependent modelling bias. This bias would strengthen, rather than erase, the anisotropy evolution inferred with a redshift-independent bias correction. However, because the radial anisotropy profile and its evolution with redshift remain poorly constrained, the redshift evolution of the modelling bias cannot yet be robustly quantified and will require more detailed investigation in future work.

Another potential source of systematic uncertainty is the use of the lens light distribution as a proxy for the stellar mass distribution. In our analysis, the lens light distribution enters the modelling in three ways: it determines the inner density profile and the prior range on the axis ratio of the BPL model, represents the tracer density in the spherical Jeans modelling, and is used to estimate the projection bias through its axis ratio. Although the HST F814W light distribution provides a reasonable proxy for the stellar mass distribution of the lenses considered here, variations in the stellar mass-to-light ratio with waveband, redshift or radius could make the light distribution a biased tracer of the stellar mass distribution and introduce systematics into the lens modelling, dynamical modelling and projection-bias estimation. These effects should be quantified and assessed using high-resolution hydrodynamical simulations, deeper multi-band imaging and spatially resolved spectroscopy across the relevant redshift range. 

Future lens samples will make this framework substantially more powerful. Increasing the number of galaxy-scale lenses will reduce the statistical uncertainties on $\gamma_{\rm PPN}$ and the anisotropy-evolution parameters, while also improving sensitivity to population-level parameters associated with external convergence and to cosmological parameters entering distance ratios. Constraints on the line-of-sight density-contrast parameters may help diagnose whether an apparent redshift dependence is associated with lens selection, lens environment, or evolution in the density field sampled by strong lenses. This connection to redshift-survey information could be strengthened by noting that spectroscopic environments and large-scale-structure maps can provide external priors on $\kappa_{\rm env}$ and $\kappa_{\rm LSS}$. Furthermore, spatially resolved kinematics for an increasing subset of lenses would provide stronger constraints on their stellar orbital anisotropy and help reduce their respective $X$--$\beta$ degeneracies. Together, larger samples of galaxy-scale lenses and more precise lens and dynamical modelling can make this framework a precise probe of gravity, orbital-structure evolution and cosmological parameters, while helping to constrain the assembly history and feedback mechanisms of massive galaxies.

\section{Methods}\label{sec11}

\subsection*{Lens sample}
The lens sample for this study is drawn from a compilation of 125 strong lensing systems, dominated by single-deflector systems and with clear lensing features. This catalogue comprises 63 lenses from the Sloan Lens ACS Survey (SLACS) \cite{2008ApJ...682..964B}, 38 from SLACS for the Masses Survey (S4TM) \cite{2017ApJ...851...48S}, and 24 from the Baryon Oscillation Spectroscopy Survey (BOSS) Emission-Line Lens Survey (BELLS) \cite{2012ApJ...744...41B}. All lenses have been imaged with the Advanced Camera for Surveys (ACS) aboard the Hubble Space Telescope (HST) through the F814W filter, with a pixel scale of $0\farcs05$. To eliminate systems with potentially unreliable velocity dispersion measurements or lens mass reconstructions, we calculated the single-aperture velocity dispersion of each lens using the best-fitting lens mass distributions and assuming isotropic stellar orbits. We then compared these predictions to the velocity dispersions reported in the main spectroscopic catalogue from the SDSS DR17 \cite{2022ApJS..259...35A}. We identified four outlier systems, all from the high-redshift BELLS sample, for which the observed velocity dispersions differ from the model-predicted values by more than 100 km/s. Among these, three had failed velocity dispersion fits in DR17  (two reached the velocity fitting boundaries and one returned an unphysical value exceeding 700 km/s), while the fourth showed an estimated value approximately 90\% higher than the corresponding SDSS measurement. By excluding these four outliers, our final sample comprises 121 lens systems for subsequent hierarchical Bayesian inferences. This selection provides a homogeneous data set in the F814W bandpass, with broadly similar spectroscopic selection effects, thereby facilitating consistent L\&D analyses across the different surveys. The properties of the compiled lens systems are summarized in Table~\ref{tabA3:lens_sample}.

\subsection*{Lens modelling}
We perform the lensed image reconstruction following the two-step procedure validated in ref. \cite{2023ApJ...953..189D}. We first fit the foreground lens light distribution using a B-spline technique with even multipoles, including additional odd terms when obvious odd features remain in the residuals after the subtraction of even terms. After subtracting the B-spline fits from observed images, we reconstruct the residual lensed images for each lens system by forward modelling, in which an elliptical broken power-law (BPL) model and an S\'ersic profile are adopted to model the lens mass and source light distributions, respectively. The source light is mapped to the image plane through the lens equation, convolved with the point spread function, and compared with the extracted lensed image. We define the posterior distribution using the likelihood and physically motivated priors, and sample it with the Python module \texttt{emcee} \cite{2013PASP..125..306F}, an affine-invariant Markov chain Monte Carlo (MCMC) ensemble sampler.

The BPL model has four radial parameters: the inner slope $\alpha_c$, the break radius $r_c$, the outer slope $\alpha$ and the amplitude parameter $b$. It can accommodate a wide range of density profiles, from flat cores to steep cusps, and can reduce to the single power-law model when the inner and outer slopes are identical. This flexibility allows the model to capture radial variations in the mass distribution across the regions probed by lensing and stellar dynamics. The theoretical formulation of the BPL model is given in ref. \cite{2020ApJ...892...62D}, and its performance in lens mass modelling has been validated in ref. \cite{2023ApJ...953..189D} through end-to-end tests, ranging from the generation of mock images to the prediction of velocity dispersions.

Physically motivated priors are essential for the robust application of the BPL model. Strong lensing is generally unable to probe the central mass distribution of the lens, especially when the central region lacks lensing signals or is significantly contaminated by the foreground lens light. We therefore regularize the inner density profile of the BPL model by fixing the inner slope $\alpha_{\rm c}$ and break radius $r_{\rm c}$ in the lens mass modelling. These two parameters are determined in advance from a fit to the lens light distribution using a power-law S\'ersic profile, whose three-dimensional density follows a power law with slope $\alpha_{\rm c}$ within the break radius $r_{\rm c}$ and transitions to a de-projected S\'ersic profile at larger radii, assuming a constant stellar mass-to-light ratio. This choice is motivated by the expectation that baryons dominate the mass budget in the central regions of massive galaxies. Consequently, we have six free parameters for each BPL lens model: two radial parameters $\alpha$ and $b$, two centroid coordinates, a position angle and an axis ratio. The outer slope is restricted to $1.8<\alpha<2.2$. We also add an empirical prior on the axis ratio of the lens mass distribution, informed by its correlation with the stellar mass distribution. With this prescription, Ref. \cite{2023ApJ...953..189D} showed that the projected mass distribution can be recovered reliably, with little dependence on exposure time and with median biases of less than 5\% for the mean convergence profile out to several Einstein radii.

We first fit all 125 lenses with a single background source. For systems with significant residuals in the reconstructed lensed images, we add a second S\'ersic source and repeat the modelling. In total, 25 systems require two background sources. We do not introduce additional source components to further improve the image reconstruction, as our primary goal is to constrain the lens mass distribution, which benefits only marginally from additional source components. In the L\&D analysis, we estimate the joint probability distribution $P(\alpha,b)$ on a $101\times101$ regular grid by assigning the MCMC posterior samples of $\alpha$ and $b$ to this grid using the Triangular Shaped Cloud scheme. This allows us to propagate the uncertainties in the lens mass distribution to the predicted stellar velocity dispersion (see Eq. \ref{eq:likelihood}).

\subsection*{Dynamical modelling}\label{subsec:dyn_model}
We predict the aperture- and luminosity-weighted LOS velocity dispersion from the reconstructed lens radial mass distribution using spherical Jeans modelling \cite{2008gady.book.....B}. For each lens, we assume that the tracer number density profile follows the stellar luminosity density profile inferred from the lens light distribution in the HST F814W band. The aperture weighting is modeled by a Gaussian function that approximates the combined effects of the spectroscopic aperture and atmospheric seeing for each lens. The fibre diameter is $3^{\prime\prime}$ for the original SDSS spectrographs and $2^{\prime\prime}$ for the BOSS spectrographs. 

We assume a constant stellar orbital anisotropy for each lens in the Jeans dynamical modelling. We interpret this parameter as a stellar-mass-weighted effective global anisotropy, following the definition \cite{2007MNRAS.379..418C,2017MNRAS.469.1824X},
\begin{equation}
\beta = 1-\frac{\Pi_{\theta\theta}+\Pi_{\phi\phi}}{2\Pi_{rr}},
\end{equation}
where $\Pi_{kk}=\sum_i M_i\sigma^2_{k,i}$ measures twice the kinetic energy in random stellar motions along the $k$-direction in spherical coordinates, with $M_i$ and $\sigma_{k,i}$ being the total stellar mass and velocity dispersion in the $i$-th radial bin, respectively. By definition, $\beta$ ranges from $-\infty$ to $1$, with $\beta=0$ for isotropic orbits, and $\beta>0~(\beta<0)$ for radially (tangentially) biased orbits. The anisotropy constrained in this work thus represents an effective global property of the stellar orbits, rather than the local anisotropy at a particular radius.

Given the estimated two-dimensional lens mass and stellar light density profiles alongside a constant $\beta$, we can compute the model-predicted LOS velocity dispersion $\sigma_{\parallel,\rm len}$ by accounting for the fibre size and seeing effects \cite{2020ApJ...892...62D,2023ApJ...953..189D}. Note that the lens mass distribution inferred from strong lensing assumes zero external convergence by default; the true mass amplitude is therefore rescaled by a factor of $1-\kappa_{\rm ext}$, where $\kappa_{\rm ext}$ is the unknown actual external convergence. In addition, deprojecting the two-dimensional lens mass and light distributions under the assumption of spherical symmetry may introduce a projection bias $b_\sigma$ in the predicted velocity dispersion. For BPL-based models, this projection bias can be corrected empirically for each lens by the relation $b_\sigma = 1.015q_{\star}^{-0.07}$ \cite{2020ApJ...892...62D}, where $q_{\star}$ is the axial ratio of the lens light distribution. Calculating $\sigma_{\parallel,\rm len}$ requires an assumption about the cosmology through the angular diameter distance ratio $D_s/D_{ds}$, where $D_s$ and $D_{ds}$ are the distances from the observer to the source and from the lens to the source, respectively. The inconsistency between the assumed and true underlying cosmology can induce an additional bias, although L\&D analyses have no direct dependence on the Hubble constant $H_0$. Potential deviations from general relativity on galaxy scales, quantified by the post-Newtonian parameter $\gamma_{\rm PPN}$ in this work, can also cause a discrepancy between $\sigma_{\parallel,\rm obs}$ and $\sigma_{\parallel,\rm len}$. Finally, the mismatch between observed and predicted velocity dispersions arises not only from stellar orbital anisotropy, but also from projection effects, external convergence, possible deviations from general relativity, and cosmological distance ratios. We therefore encapsulate these effects into an effective mismatch parameter $X$ and construct a likelihood for each lens in the reduced $X$--$\beta$ parameter space.

\subsection*{Likelihood} 
For each lens, we solve the spherical Jeans equation by assuming a global constant anisotropy parameter $\beta$. The aperture- and luminosity-weighted LOS velocity dispersion $\slen$ is calculated based on the lens galaxy's luminosity profile, measured directly from the F814W-band image, and its mass-density profile, described by the broken power-law model with four radial parameters, $\alpha_c$, $r_c$, $\alpha$, and $b$. Here, $\alpha_c$ and $r_c$ are determined from the luminosity profile fitting and are fixed in both the lens and dynamical modelling. Therefore, the probability of the observed velocity dispersion $\sobs$ from SDSS spectroscopy, given a set of specific parameters, is
\begin{equation}
  P(\sobs|X,\beta,\alpha,b)=\frac{1}{\sqrt{2\pi}\tau} \exp{\left[-\frac{(\sobs-X\slen)^2}{2\tau^2}\right]} \label{eq:psigma}
\end{equation}
where
$\tau^2=\tau_{\rm obs}^2+\tau_{\rm in}^2$ accounts for the measurement uncertainty $\tau_{\rm obs}$ and an intrinsic scatter of $\tau_{\rm in}\simeq0.06\sobs$. As described in the preceding subsection, the factor $X$ quantifies the mismatch between $\sobs$ and $\slen$ and can be written as
\begin{equation}
  X=\frac{1}{b_\sigma}\sqrt{\frac{1-\kappa_{\rm ext}}{1-\kappa_{\rm ext}^\prime}\frac{2}{1+\gamma_{\rm PPN}}\frac{\mathcal{R}}{\mathcal{R}^\prime}} ~,
\end{equation}
where $b_\sigma$ denotes the projection bias, or more generally a dynamical-modelling bias, $\kappa_{\rm ext}=\kappa_{\rm LSS}+\kappa_{\rm env}$ is the actual external convergence with $\kappa_{\rm LSS}$ and $\kappa_{\rm env}$ being the external convergence from uncorrelated large-scale structures and the local environment respectively, $\kappa_{\rm ext}^\prime$ is the external convergence assumed in the lens mass modelling, and $\gamma_{\rm PPN}$ denotes the parameterized post-Newtonian parameter. $\mathcal{R}={D_s}/{D_{ds}}$ is the distance ratio for the true cosmology, while $\mathcal{R}^\prime$ is the distance ratio for an adopted fiducial cosmology. In this paper, we assume $\kappa_{\rm ext}^\prime = 0$ in the lens mass modelling. 

For each lens, we then construct a likelihood for $(X,\beta)$ by marginalizing over the lensing-inferred parameters $(\alpha,b)$:
\begin{equation}
  \mathcal{L}(X,\beta)=P(\sobs|X,\beta)=\iint P(\sobs|X,\beta,\alpha,b)P(\alpha,b) {\rm d}\alpha{\rm d}b \label{eq:likelihood}
\end{equation}
where $P(\alpha,b)$ represents the posterior distribution of $(\alpha,b)$ derived from the MCMC samples of the lens modelling.

\subsection*{Prior on external convergence} 

For each lens, we decompose the external convergence as $\kappa_{\rm ext}=\kappa_{\rm env}+\kappa_{\rm LSS}$ and adopt a Gaussian prior for $\kappa_{\rm ext}$ centred on $\kappa_{\rm env}$ with variance $\langle\kappa_{\rm LSS}^2\rangle$. Here, $\kappa_{\rm env}$ accounts for the local lens environment, whereas $\kappa_{\rm LSS}$ represents the contribution from uncorrelated large-scale structure along the line of sight.

We first identify lenses projected within the radius $r_{200}$ of a massive cluster, where $r_{200}$ is the radius within which the mean density of the cluster is 200 times the critical density of the Universe at the cluster redshift. Specifically, we cross-match the lens sample with the redMaPPer cluster catalogue \cite{2016ApJS..224....1R} and find that 13 lenses are member galaxies of redMaPPer clusters, five of which are central galaxies. For these central cluster galaxies, we set $\kappa_{\rm env}=0$ to avoid double-counting the lens mass. For the remaining eight satellite galaxies, we estimate $\kappa_{\rm env}$ as the convergence contributed by the host cluster at the projected cluster-centric radius of each lens. We model each cluster with an NFW profile. The halo mass is inferred from its redMaPPer richness via the redMaPPer mass--richness relation, and the corresponding concentration is obtained using the mass--concentration relation derived from CFHTLenS clusters \cite{2015ApJ...814..120D}. These two quantities specify the cluster mass profile, from which we compute the convergence at the lens position. In addition, three lenses are projected near redMaPPer clusters but have large redshift offsets from the clusters. Their cluster-induced external convergence is found to be negligible ($<0.005$), and these structures are treated as part of the uncorrelated large-scale structure along the line of sight. Details of the lens--cluster associations, including $\kappa_{\rm env}$, are provided in Table~\ref{tabA2:redmapper}.

The weak lensing effects from LSS along the LOS can bias the predicted velocity dispersion squared by a factor, denoted as $1-\kappa_{\rm LSS}$. Equivalently, this term describes the multiplicative contribution of uncorrelated foreground and background structure to the lensing-dynamical normalization. Due to the non-linear coupling between main lens and LOS structures, this factor still depends on the actual lens mass and light distributions and stellar velocity anisotropy \cite{2017JCAP...04..049B}. For a lens galaxy with spherical power-law mass and light density profiles, and a constant velocity anisotropy, this factor can be written as \cite{2022JCAP...07..027T,2024JCAP...10..055J}
\begin{equation}
1-\kappa_{\rm LSS}=\frac{1-\kappa_s}{1-\kappa_{ds}}
\end{equation}
where $\kappa_s$ and $\kappa_{ds}$ are the weak lensing convergences evaluated from the observer to the source and from the lens to the source, respectively. We use this relation as a first-order approximation to estimate $\kappa_{\rm LSS}$ for the BPL-based models considered here. In the weak lensing limit, we further have 
\begin{equation}
\kappa_{\rm LSS}\simeq (\kappa_s-\kappa_{ds})=\int_0^{\chi_s} d\chi ~(1-\mathcal{B})q(\chi)\delta(\chi\boldsymbol{\theta},\chi) \label{eq:klss}
\end{equation}
with $\chi$ being the comoving radial distance for a flat Universe and 
\begin{equation}
\mathcal{B}=\frac{\chi_s}{\chi}\frac{\chi-\chi_d}{\chi_s-\chi_d}
\end{equation}
for $\chi > \chi_d$, and $\mathcal{B}=0$ otherwise. Here, $\chi_d$ and $\chi_s$ denote the comoving distances from the observer to the lens and the source, respectively. $\delta(\chi\boldsymbol{\theta},\chi)$ is the matter density contrast at distance $\chi$ and angular position $\boldsymbol{\theta}$. The weak lensing weight function $q(\chi)$ is defined by
\begin{equation}
q(\chi)=\frac{3H_0^2\Omega_m}{2c^2}\frac{\chi}{a(\chi)}\frac{\chi_s-\chi}{\chi_s}
\end{equation}
 
Starting from Eq. \ref{eq:klss}, the variance of external convergence from LSS smoothed with a top-hat window $U(\theta_{\rm ap})$ of radius $\theta_{\rm ap}$ can be calculated by 
\begin{equation}
\langle\kappa_{\rm LSS}^2(\theta_{\rm ap})\rangle=\int\frac{d^2\ell}{(2\pi)^2}P_\kappa(\ell)|\hat{U}(\ell)|^2
\end{equation}
where $\hat{U}(\ell)$ is the Fourier transform of $U(\theta_{\rm ap})$ and $P_\kappa(\ell)$ is the external convergence power spectrum defined as
\begin{equation}
P_\kappa(\ell)=\int d\chi \frac{q^2(\chi)}{\chi^2}P_\delta(\frac{\ell}{\chi},\chi) (1-\mathcal{B})^2
\end{equation}

In this work, we adopt the Einstein radius $\theta_E$ of each lens system as the aperture radius to evaluate $\langle\kappa_{\rm LSS}^2(\theta_E)\rangle$, with the non-linear matter power spectrum computed using CAMB \cite{2012JCAP...04..027H} for the fiducial Planck cosmology \cite{2020A&A...641A...6P}.

\subsection*{Hierarchical Bayesian framework} 
To facilitate efficient posterior sampling, we pre-calculate the likelihood function $P(\sobs|X,\beta)$ on a $128\times128$ grid over the parameter space $(X,\beta)$, with $X$ ranging from 0.7 to 1.3 and $\beta$ from -2 to 1. The range of $X$ is primarily determined by $\kappa_{\rm ext}\in(-0.1,0.4)$, $\gamma_{\rm PPN}\in(0.5,1.5)$, and $b_\sigma\in(1.0,1.1)$, and is sufficiently wide to accommodate additional uncertainties in the cosmological distance ratio $\mathcal{R}$. We parameterize the velocity anisotropy as $\beta(z)=\beta_0+\beta_zz$ with an intrinsic scatter of $\tau_\beta$, and the average trend of density contrast along the line-of-sight towards strong lensing systems as $\delta_m(z)=\delta_{m,0}+\delta_{m,z}z$. 

Based on the parametrization of $\delta_m(z)$, we can calculate the $\kappa_{\rm LSS}$ according to Eq. \ref{eq:klss}. In practice, for a lens at redshift $z_{d}$ with a source at $z_{s}$, the effective external convergence from uncorrelated large scale structure is estimated by $\kappa_{\rm LSS}=\sum_iw_i(z_d,z_s,z_i)\delta_m(z_i)$ with $z_i$ denoting the redshift of the $i$-th redshift bin between observer and source. In the discrete case, $w_i$ can be interpreted as
\begin{equation}
w_i(z_d,z_s,z_i)=[1-\mathcal{B}(z_d,z_s,z_i)]\frac{\bar{\rho}_m(z_i)}{\Sigma_{\rm crit}(z_i,z_s)}\Delta D_{p,i}
\end{equation}
Here, $\bar{\rho}_m(z_i)$ is the mean matter density of the Universe at redshift $z_i$, $\Delta D_{p,i}$ is the proper radial distance interval, set to $50~{\rm Mpc}$, and $\Sigma_{\rm crit}$ is the critical surface mass density. 

In a hierarchical Bayesian framework, the marginal likelihood for each lens given the population-level parameters is obtained by marginalizing over its lens-specific orbital anisotropy. For lens $j$, this gives 
\begin{align}
&P(\sobsj|\gamma_{\rm PPN},\beta_0,\beta_z,\tau_\beta,\delta_{m,0},\delta_{m,z}) = \notag \\
&\int P[\sobsj|X^j(\gamma_{\rm PPN},\delta_{m,0},\delta_{m,z}),\beta^j]P(\beta^j|\beta_0,\beta_z,\tau_\beta) \,d\beta^j ,
\end{align}
where $P(\beta^j|\beta_0,\beta_z,\tau_\beta)$ serves as a population-level prior for the anisotropy of lens $j$ at its redshift. Thus, the log-posterior for $N$ strong lensing systems can be written as
\begin{align}
&\ln P(\gamma_{\rm PPN},\beta_0,\beta_z,\tau_\beta,\delta_{m,0},\delta_{m,z}|\{\sobsj\})= \notag \\
&\sum_{j=1}^N\ln P(\sobsj|\gamma_{\rm PPN},\beta_0,\beta_z,\tau_\beta,\delta_{m,0},\delta_{m,z})+\lambda\sum_{j=1}^N\ln P(\kappa_{\rm ext}^j)+ \mathrm{const.} \label{eq:bayes}
\end{align}
Here, $P(\kappa_{\rm ext}^j)$ denotes the Gaussian prior on the external convergence of lens $j$, with mean $\kappa_{\rm env}^j$ and standard deviation $\langle(\kappa_{\rm LSS}^j)^2\rangle^{1/2}$, evaluated at the value of $\kappa_{\rm ext}^j$ implied by the sampled values of $\delta_{m,0}$ and $\delta_{m,z}$. For sensitivity analysis, we introduce a scaling factor $\lambda$ that controls the weight assigned to the external-convergence prior, with $\lambda=1$ corresponding to the standard Bayesian case and $\lambda<1$ down-weighting the external convergence prior. Varying $\lambda$ allows us to assess the sensitivity of the parameter inference to external convergence. Uniform priors are applied to the population-level parameters. Specifically, we impose the physical constraints $\delta_m(z)>-1$ and $\beta(z)<1$ at all redshifts. We sample the joint posterior of the population-level parameters using \texttt{emcee}.

\subsection*{Literature estimates of stellar orbital anisotropy}
To contextualize our constraints on the redshift evolution of stellar orbital anisotropy $\beta(z)=\beta_0+\beta_z z$, we compare our result with literature estimates of $\beta$ for massive galaxies from hydrodynamical simulations and observations. For the simulation comparisons, we use the dynamical properties of massive galaxies from the Illustris-1 and TNG100 cosmological hydrodynamical simulations \cite{2017MNRAS.469.1824X,2026ApJ...998..303L}. For Illustris-1, we analyze the public catalogue from ref.~\cite{2017MNRAS.469.1824X} and select galaxies over $0.1<z<0.7$ with luminosity-weighted LOS velocity dispersions of $160$--$400~{\rm km/s}$, measured within half an effective radius. This velocity dispersion range follows that adopted in ref.~\cite{2017MNRAS.469.1824X}. We then use the stellar-mass-weighted $\beta$ estimated within a three-dimensional radius corresponding to an aperture of $1.5^{\prime\prime}$ from the galaxy centre, comparable to the median effective radius of our lens sample and to the SDSS spectroscopic fibre radius. A linear fit to the resulting $\beta$--$z$ relation gives an intercept of $0.17$ and a slope of $-0.36$. For TNG100, we use the galaxy catalogue provided by the authors of ref.~\cite{2026ApJ...998..303L}, applying the same velocity-dispersion selection over $0.01<z<0.7$. In this case, $\beta$ is the stellar-mass-weighted anisotropy measured within one effective radius, and the linear fit gives an intercept of $0.27$ and a slope of $-0.01$.

For observational comparisons, we compile published $\beta$ estimates for local galaxies and for lens samples analysed in previous L\&D studies. For local galaxies, we adopt the spherical Jeans anisotropic modelling (JAM) results for galaxies from the MaNGA integral-field spectroscopic survey \cite{2023MNRAS.522.6326Z}. We select 371 galaxies with velocity dispersions of $160$--$400~{\rm km\,s^{-1}}$ and the highest modelling quality, \texttt{qual=3}. This sample has a median redshift of $z\simeq0.05$, and yields median anisotropies of $\beta = 0.47^{+0.13}_{-0.27}$ for the JAM model with a fixed NFW halo and $\beta = 0.24^{+0.27}_{-0.44}$ when the characteristic density of the NFW halo is left free.

Large-sample L\&D analyses have commonly adopted a single power-law mass model, while the stellar orbital anisotropy has often been fixed to isotropy or constrained with external Gaussian priors, such as a prior centred on $\beta=0.18$ with a standard deviation of $0.13$ \cite{2006PhRvD..74f1501B,2010ApJ...708..750S,2019MNRAS.488.3745C}. Direct lens-sample constraints on $\beta$ are therefore limited. For this comparison, we include the ensemble-averaged measurement of $\beta$ from 58 SLACS lenses with median $z\simeq0.19$ by ref.~\cite{2009ApJ...703L..51K}. By calibrating the lens mass-density slope using dynamics-independent scaling relations, they derived an average orbital anisotropy of $\beta=0.45\pm0.25$. We also include the population-level constraints from Project Dinos, which aimed to constrain the redshift evolution of mass-density profile in elliptical galaxies \cite{2024MNRAS.530.1474T,2025MNRAS.541....1S}. Dinos I \cite{2024MNRAS.530.1474T}, which focused on the deviation of the total mass-density profile from a power-law, inferred a population mean of $\beta = 0.53^{+0.27}_{-0.45}$ for 77 lenses (34 SLACS, 24 SL2S and 19 BELLS) with median $z\simeq0.41$. Dinos II \cite{2025MNRAS.541....1S}, which focused on the population-level properties of the dark and luminous matter distributions, obtained $\beta = -0.04^{+0.19}_{-0.22}$ for 33 SLACS lenses with median $z\simeq0.2$ and $\beta = -0.19^{+0.19}_{-0.19}$ for 23 SL2S lenses with median $z\simeq0.55$.

Across these literature studies, the galaxy mass ranges are broadly comparable. However, the data sets and modelling methodologies differ substantially. We therefore use these results only for comparison in Fig.~\ref{fig:beta_z}, without homogenizing them with our modelling assumptions.

\backmatter



\section*{Acknowledgements}
We thank Dandan Xu and Yan Liang for providing and interpreting measurements of stellar orbital anisotropy in TNG100 galaxies, and for insightful discussions of its redshift evolution. The authors also acknowledge Beijing PARATERA Tech CO., Ltd. (\url{https://www.paratera.com/}) for providing HPC resources that have contributed to the research results reported within this paper.

\section*{Funding}
W.D. and L.P.F. are supported by the National Natural Science Foundation of China (NSFC) under grant no. 12541302. 
L.P.F. is also supported by the Innovation Program of the Shanghai Municipal Education Commission under grant no. 2025GDZKZD04. 
G.B.Z. is supported by the National Key R\&D Program of China (2023YFA1607800, 2023YFA1607803), the NSFC grant 12525301, and the New Cornerstone Science Foundation through the XPLORER prize.
G.B.Z. and S.Y. are supported by the CAS Project for Young Scientists in Basic Research under grant no. YSBR-092 and the China Manned Space (CMS) Program under grant no. CMS-CSST-2021-B01.
S.Y. is also supported by the NSFC under grant no. 12203062.
Z.H.F. is supported by the CMS Program under grant no CMS-CSST-2021-A01 and the NSFC under grant no. 11933002.
H.Y.S. is supported from the NSFC under grant no. 12533008.
This work is also supported by the CMS Program under grant nos. CMS-CSST-2025-A02, CMS-CSST-2025-A03, CMS-CSST-2025-A05 and CMS-CSST-2025-A20.

\section*{Competing interests}
The authors declare no competing interests.

\bibliography{sn-bibliography}

\begin{appendices}
\newpage

\newgeometry{top=1.in,bottom=1.5in}
\section{Supplementary information}
\vspace{\baselineskip}

\subsection{Examples of lens modelling}\label{secA1:lens_modelling}
\begin{figure}[!ht]
    \centering
    \includegraphics[width=\linewidth]{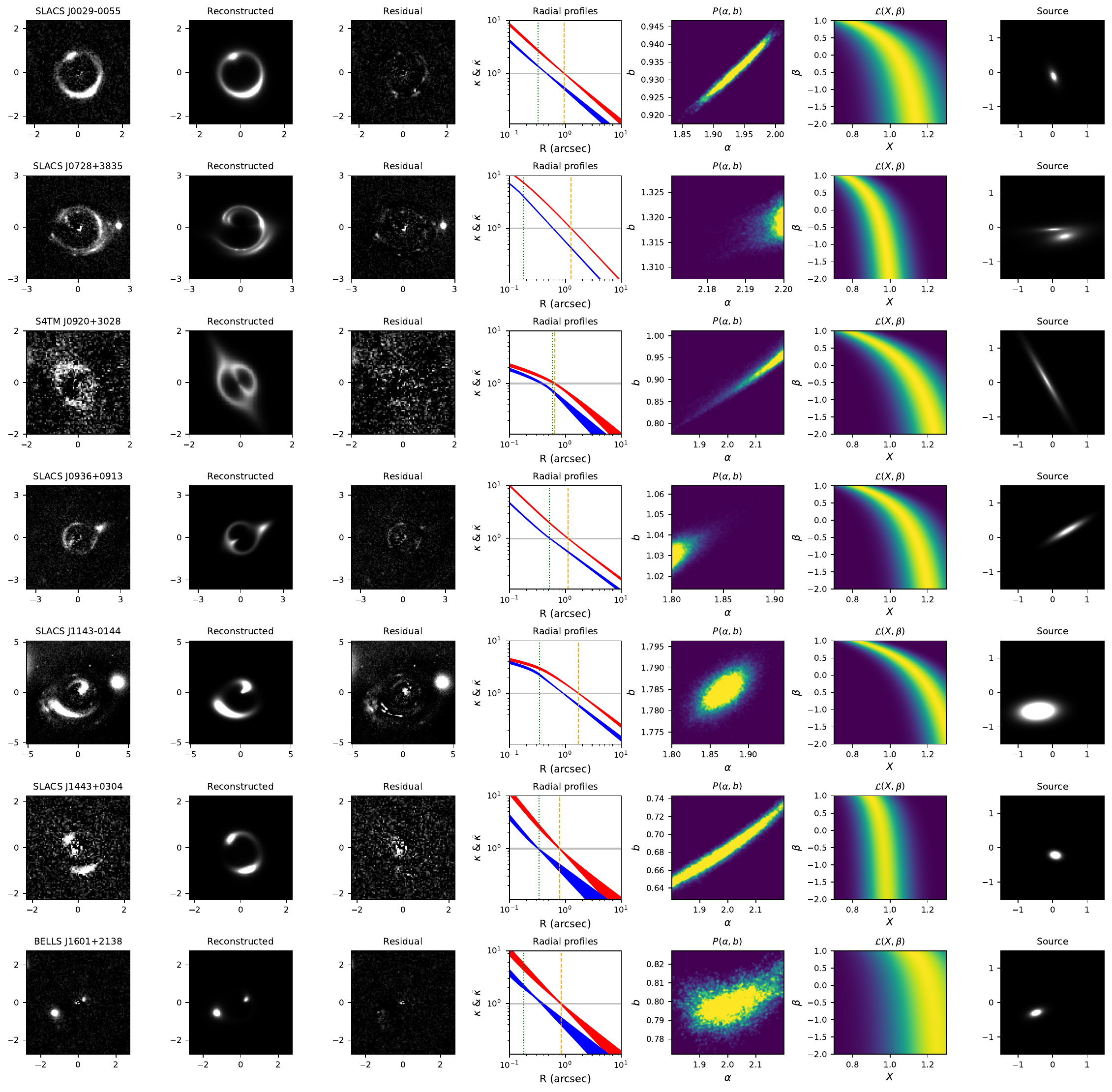}
    \caption{Examples of lens modelling. We show the lens modelling results for seven representative lenses. From left to right, the first three columns show the lensed images with lens light subtracted, the reconstructed images and the residuals, respectively. The fourth column shows the reconstructed radial profiles of the convergence (blue bands) and mean convergence (red bands), with the bands indicating the 95\% percentile intervals. The vertical green dotted and orange dashed lines in each panel indicate the break radius and Einstein radius, respectively. The fifth and sixth columns show the posterior distribution $P(\alpha,b)$ and the inferred likelihood $\mathcal{L}(X,\beta)$, respectively, illustrating the degeneracies between the mass-profile parameters and between the effective mismatch parameter $X$ and stellar orbital anisotropy $\beta$. The last column shows the reconstructed source images.}
    \label{figA1:bpl_fitting}
\end{figure}

\newpage
\subsection{Sensitivity tests}\label{secA2:sensitivity_test}
\begin{figure}[H]
    \centering
    \includegraphics[width=\linewidth]{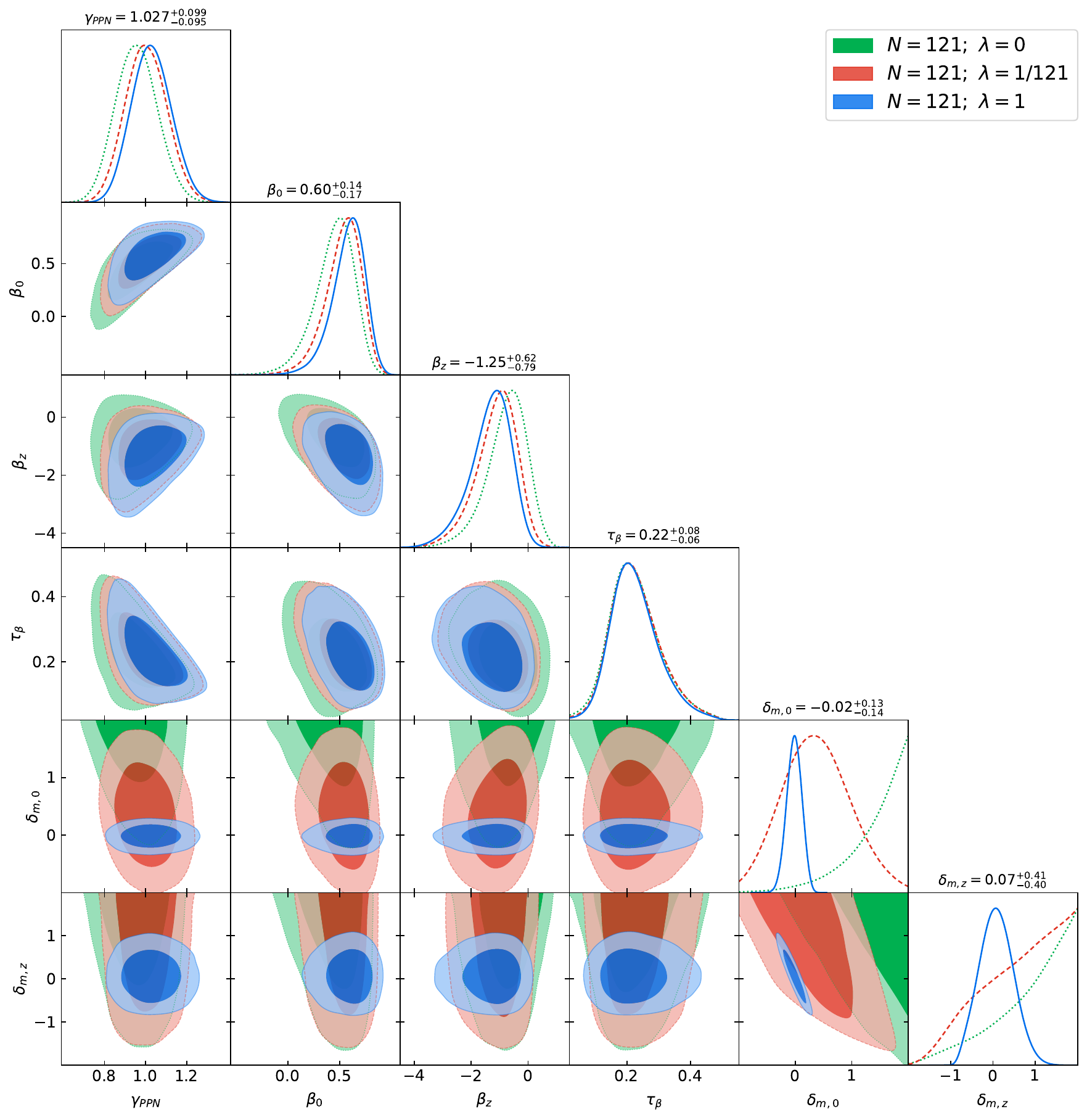} 
    \caption{Posterior distributions of population parameters from 121 galaxy-scale lenses (63 SLACS, 38 S4TM and 20 BELLS). Diagonal panels show the one-dimensional marginalized posterior distributions, with the median and 68\% credible interval for $\lambda=1$ reported above each panel. Off-diagonal panels show the two-dimensional posteriors, with contours enclosing the 68\% and 95\% credible regions. The green, red and blue posterior distributions correspond to Bayesian inference with $\lambda=0$, $1/121$ and $1$, respectively. The results demonstrate that the posterior distributions of $\gamma_{\rm PPN}$, $\beta_0$, $\beta_z$ and $\tau_\beta$ depend only weakly on the external-convergence prior over the considered range of $\delta_{m,0}$ and $\delta_{m,z}$.}
    \label{figA2:mcmc_lam}
\end{figure}

\begin{figure}[!ht]
    \centering
    \includegraphics[width=\linewidth]{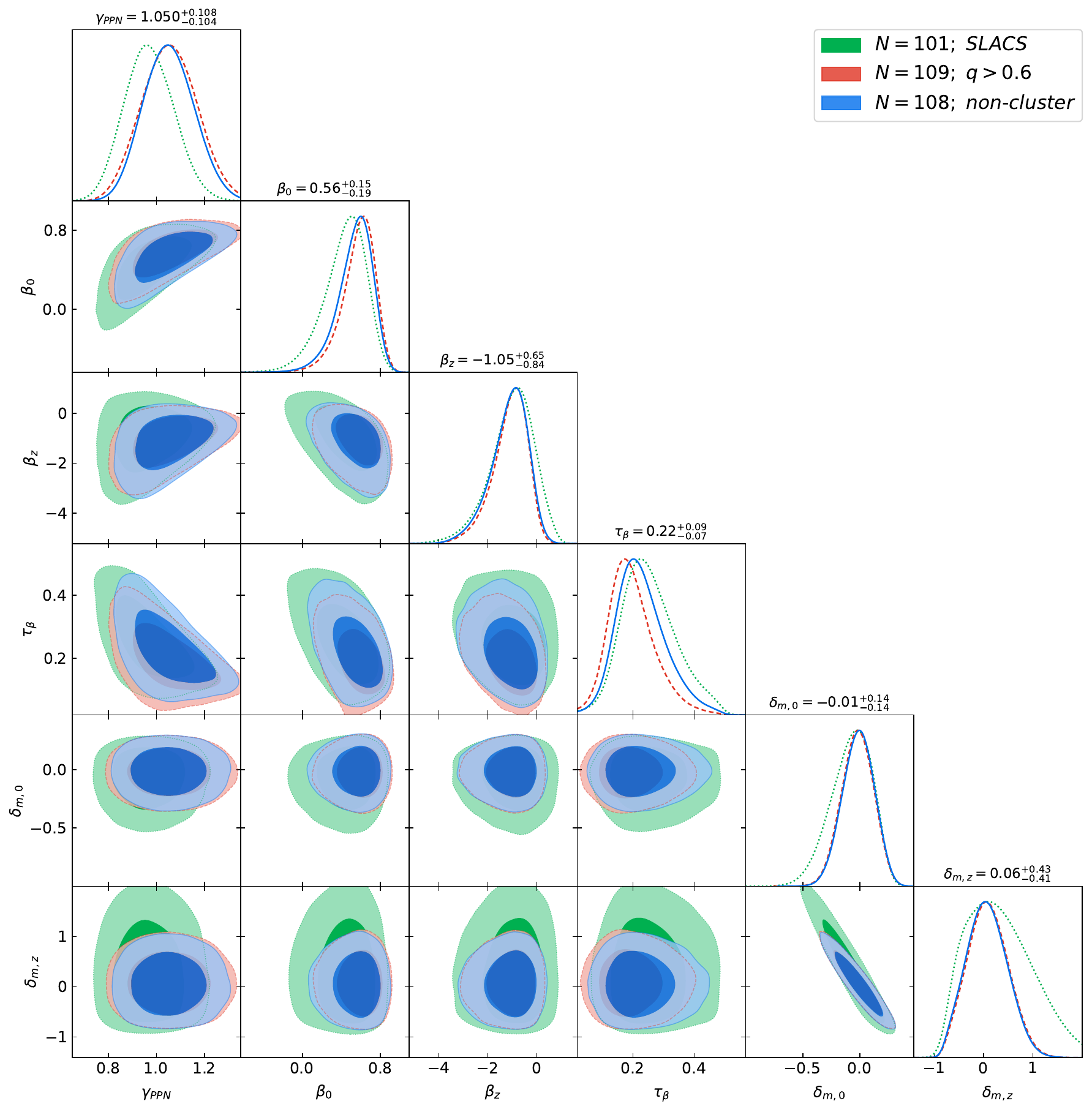}
    \caption{Similar to Fig.~\ref{figA2:mcmc_lam}, but with $\lambda=1$ and for the subsamples of 101 SLACS lenses (green), 109 lenses with projected axis ratio $q>0.6$ (red), and 108 lenses that are not members of redMaPPer clusters (blue). The inferred values of $\gamma_{\rm PPN}$, $\beta_0$, $\beta_z$ and $\tau_\beta$ are mutually consistent within their statistical uncertainties. All three subsamples remain consistent with $\gamma_{\rm PPN}=1$ and favour a positive anisotropy at low redshift together with a negative redshift dependence for the lens-galaxy population.} \label{figA3:mcmc_sam}
\end{figure}

In this section, we present tests of the sensitivity of the population-level constraints to the adopted external-convergence prior, subsample selection, background cosmology and projection bias. The results are summarized in Table~\ref{tabA1:sensitivity}, with representative posterior constraints shown in Figs.~\ref{figA2:mcmc_lam} and \ref{figA3:mcmc_sam}. These tests indicate that the main population-level constraints are not significantly affected by the adopted external-convergence prior, subsample selection or background cosmology. Setting $b_\sigma=1$ shifts the inferred $\gamma_{\rm PPN}$ upward to $1.20\pm0.11$, while leaving the anisotropy constraints broadly consistent. This indicates that the normalization of the projection bias primarily affects the absolute calibration of $\gamma_{\rm PPN}$, whereas the evolutionary trend of $\beta$ is relatively insensitive to this normalization.

\restoregeometry

\newgeometry{margin=0.5in,bottom=1.5in}
\begin{table}
\centering
\footnotesize
\renewcommand{\arraystretch}{1.5}
\caption{Sensitivity of posterior parameters to $\lambda$, sample selection, cosmology and the projection bias. Unless otherwise stated, the fiducial projection bias $b_\sigma=1.015q^{-0.07}_\star$ and fiducial Planck cosmology are adopted.}\label{tabA1:sensitivity}
\begin{tabular*}{\textwidth}{@{\extracolsep\fill}lllllll}
\toprule%
$\lambda$ & $\gamma_{\rm PPN}$(0.5,1.5)  & $\beta_{0}$(-1,1) &$\beta_{z}$(-5,2) &$\tau_{\beta}$(0.02,0.5) & $\delta_{m,0}$(-1,2) &$\delta_{m,z}$(-2,2)\\
\midrule
1          & $1.027^{+0.099}_{-0.095}$ (1.087) & $0.60^{+0.14}_{-0.17}$ (0.64) & $-1.25^{+0.62}_{-0.79}$ ($-1.03$) & $0.22^{+0.08}_{-0.06}$ (0.19) & $-0.02^{+0.13}_{-0.14}$ ($-0.01$) & $0.07^{+0.41}_{-0.40}$ (0.06) \\
1/10       & $1.025^{+0.101}_{-0.096}$ (1.086) & $0.59^{+0.14}_{-0.17}$ (0.64) & $-1.21^{+0.62}_{-0.81}$ ($-1.00$) & $0.22^{+0.08}_{-0.06}$ (0.19) & $-0.13^{+0.31}_{-0.29}$ ($-0.14$) & $0.59^{+0.86}_{-0.88}$ (0.62) \\
1/121      & $1.001^{+0.102}_{-0.099}$ (1.064) & $0.55^{+0.15}_{-0.19}$ (0.60) & $-1.04^{+0.62}_{-0.78}$ ($-0.84$) & $0.22^{+0.08}_{-0.07}$ (0.20) & \hspace{0.22cm}$0.36^{+0.62}_{-0.58}$ ($-0.10$)  & $0.80^{+0.86}_{-1.20}$ (2.00) \\
0          & $0.955^{+0.103}_{-0.100}$ (0.955) & $0.46^{+0.17}_{-0.21}$ (0.43) & $-0.68^{+0.61}_{-0.76}$ ($-0.24$) & $0.22^{+0.09}_{-0.07}$ (0.21) & \hspace{0.22cm}$1.55^{+0.33}_{-0.65}$ (2.00)         & $1.13^{+0.64}_{-1.18}$ (2.00) \\
\midrule
1 (SLACS)  & $0.967^{+0.106}_{-0.100}$ (1.027) & $0.46^{+0.18}_{-0.24}$ (0.53) & $-0.94^{+0.79}_{-0.97}$ ($-0.76$)  & $0.25^{+0.10}_{-0.08}$ (0.22) & $-0.06^{+0.17}_{-0.20}$ ($-0.04$) & $0.27^{+0.79}_{-0.64}$ (0.17)   \\
1 ($q>0.6$)  & $1.054^{+0.116}_{-0.115}$ (1.146) & $0.59^{+0.14}_{-0.18}$ (0.66) & $-1.02^{+0.60}_{-0.79}$ ($-0.75$) & $0.19^{+0.08}_{-0.06}$ (0.16) & $-0.02^{+0.14}_{-0.14}$ ($-0.02$) & $0.08^{+0.43}_{-0.41}$ (0.07)   \\
1 (non-cluster)  & $1.050^{+0.108}_{-0.104}$ (1.128) & $0.56^{+0.15}_{-0.19}$ (0.62) & $-1.05^{+0.65}_{-0.84}$ ($-0.78$) & $0.22^{+0.09}_{-0.07}$ (0.19) & $-0.01^{+0.14}_{-0.14}$ ($-0.01$) & $0.06^{+0.43}_{-0.41}$ (0.05)   \\
\midrule
1 (WMAP~9)  & $1.018^{+0.101}_{-0.097}$ (1.085) & $0.58^{+0.14}_{-0.17}$ (0.64) & $-1.18^{+0.61}_{-0.79}$ ($-0.95$)  & $0.22^{+0.08}_{-0.07}$ (0.19) & $-0.02^{+0.13}_{-0.13}$ ($-0.01$) & $0.07^{+0.40}_{-0.40}$ (0.05)   \\
1 ($\Omega_m$=0.35)  & $1.039^{+0.099}_{-0.097}$ (1.105) & $0.61^{+0.14}_{-0.17}$ (0.66) & $-1.32^{+0.63}_{-0.80}$ ($-1.06$) & $0.22^{+0.08}_{-0.06}$ (0.19) & $-0.02^{+0.13}_{-0.14}$ ($-0.02$) & $0.08^{+0.42}_{-0.41}$ (0.07)   \\
\midrule
1 ($\gamma_{\rm PPN}=1$)  & 1 & $0.56^{+0.13}_{-0.13}$ (0.56) & $-1.36^{+0.64}_{-0.72}$ ($-1.28$)  & $0.24^{+0.06}_{-0.06}$ (0.23) & $-0.01^{+0.13}_{-0.13}$ ($-0.01$) & $0.06^{+0.41}_{-0.40}$ (0.07)   \\
\midrule
1 ($b_{\sigma}=1$)  & $1.203^{+0.107}_{-0.106}$ (1.269) & $0.62^{+0.13}_{-0.16}$ (0.66) & $-1.05^{+0.56}_{-0.72}$ ($-0.87$)  & $0.19^{+0.07}_{-0.06}$ (0.17) & $-0.02^{+0.13}_{-0.13}$ ($-0.02$) & $0.07^{+0.41}_{-0.40}$ (0.07)   \\
\botrule
\end{tabular*}
\footnotetext{\textit{Notes.} Values are posterior medians with 68\% credible intervals; numbers in parentheses indicate the maximum-a-posteriori estimates. Uniform priors are used for all parameters, with ranges indicated in the table header. The prior ranges for $\delta_{m,0}$ and $\delta_{m,z}$ are motivated by inspection of the density contrast distributions at different redshifts in the L500 cosmological N-body simulation \cite{2018ApJ...853...25W}, with the range for $\delta_{m,0}$ chosen to cover approximately the 95\% highest-density interval of the density-contrast distribution in the HEALPix map with $N_{\rm side}=8192$ at redshift $z=0$. The first four rows use the full sample of 121 lenses. The next three rows report constraints for three subsamples: 101 SLACS lenses (63 SLACS + 38 S4TM), 109 lenses with projected axis ratio $q>0.6$ and 108 lenses that are not associated with redMaPPer clusters. The following two rows show results for the WMAP9 cosmology with $\Omega_m=0.28$ and a flat cosmology with $\Omega_m=0.35$. The next row shows the constraints with $\gamma_{\rm PPN}=1$. The last row reports the result for $b_{\sigma}=1$, corresponding to no correction for projection bias.} 
\end{table}
\restoregeometry

\newgeometry{top=1.5in,bottom=1.5in}
\subsection{Validation with mock dynamical data}\label{secA3:validation_test}
We validate our hierarchical inference framework using mock dynamical data generated for the same 121 lenses as in the observed sample. 
These mock data preserve the redshifts, luminosity profiles and lensing mass distributions of the observed systems, but resimulate their dynamical velocity dispersions under specified assumptions for $\gamma_{\rm PPN}$, $\beta(z)$ and external-convergence.

For each lens, we adopt the best-fitting luminosity profile as the input light distribution and use the posterior mean values of the lens mass-model parameters to define the input mass distribution. Given an assumed stellar orbital anisotropy, we compute the lensing-predicted aperture velocity dispersion, $\sigma_{\rm T,len}$, taking into account the effects of the spectroscopic aperture and atmospheric seeing. We then apply the input mismatch factor $X$ to obtain the corresponding mock ``true'' dynamical velocity dispersion, $\sigma_{\rm T,obs}$.

We construct two types of mock data. The first type includes noisy realizations and is used to test parameter recovery under realistic statistical uncertainties. For these mocks, the input population model is $\gamma_{\rm PPN}=0.9$ and $\beta(z)=0.5-0.5z$ with intrinsic scatter $\tau_\beta=0.15$. For each lens, $\beta$ is drawn from the input population distribution. The external convergence is given by a weighted sum of the density contrast along a randomly selected line of sight in the L500 simulation \cite{2018ApJ...853...25W}, where the weight is defined by Eq.~11 in the main article. We then add the actual observational uncertainty $\tau_{\rm obs}$ and an additional 6\% intrinsic scatter to $\sigma_{\rm T,obs}$ to generate the mock observed velocity dispersion $\sigma_{\parallel,\rm obs}$. When constructing the single-lens likelihood $\mathcal{L}(X,\beta)$ for these noisy mock data, we follow the same procedure as for the observed sample, in which the posterior distribution of the lens mass-model parameters, $P(\alpha,b)$, is marginalized over.

\begin{figure}[H]
    \centering
    \includegraphics[width=\linewidth]{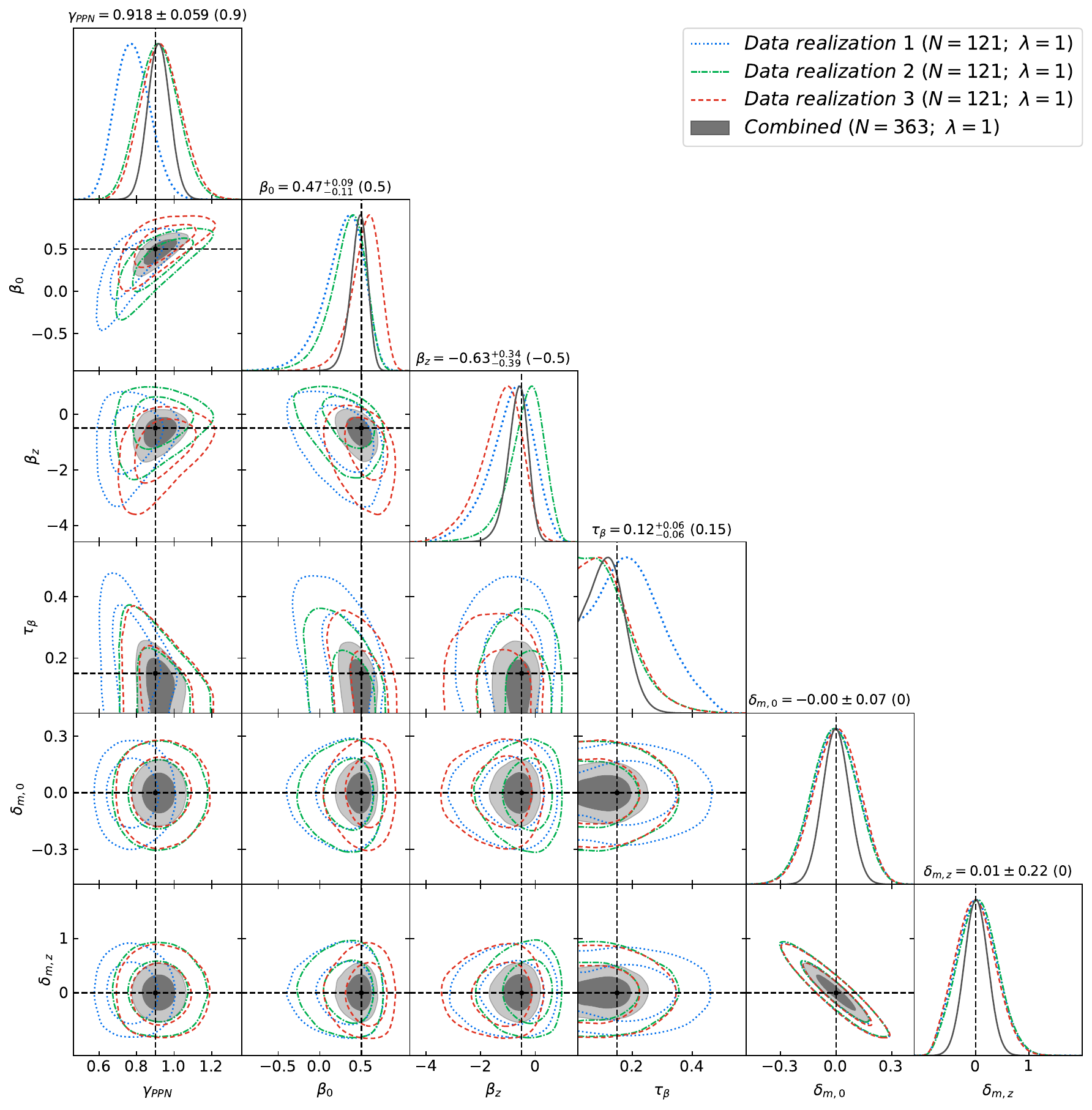}
    \caption{Recovery test with noisy mock observed velocity dispersions. The mock data are generated with input values $\gamma_{\rm PPN}=0.9$, $\beta(z)=0.5-0.5z$ and $\tau_\beta=0.15$; the external convergence of each lens is computed from a randomly selected line of sight in the L500 simulation, corresponding to $\delta_{m,0}\simeq0$ and $\delta_{m,z}\simeq0$ at the population level. The blue dotted, green dash-dotted and red dashed curves and contours show the constraints from three independent noisy realizations, while the grey curves and shaded contours show the combined constraint. The input values are indicated by the black dashed lines and by the values in parentheses above each diagonal panel. Diagonal panels show the one-dimensional marginalized posteriors, with the median and 68\% credible interval for the combined sample reported above each panel. Off-diagonal panels show the two-dimensional marginalized posteriors, with contours enclosing the 68\% and 95\% credible regions. The input parameters are recovered within the statistical uncertainties, and the combined sample gives tighter and more stable constraints.}
    \label{figA4:mcmc_sim}
\end{figure}

Fig.~\ref{figA4:mcmc_sim} shows the posterior distributions from three independent noisy realizations, each containing 121 lenses, and from their combined sample of 363 lenses. Across the three realizations, the input parameters are recovered within the statistical uncertainties, with realization-to-realization variations expected for samples of 121 lenses. The combined sample gives tighter and more stable constraints, demonstrating that the recovery becomes more stable as the sample size increases.

The second mock data set is noise-free and is used to isolate the constraining power of the inference framework. In this case, each lens is assigned the deterministic anisotropy value given by $\beta(z)=0.5-0.5z$ at the lens redshift, without intrinsic scatter in $\beta$. The external convergence is specified using $\delta_m(z)=0.3-0.3z$, and the mock observed velocity dispersion is set to be the true value, $\sigma_{\parallel,\rm obs}=\sigma_{\rm T,obs}$. When constructing the single-lens likelihoods, however, we still include the observational uncertainty $\tau_{\rm obs}$ and the intrinsic scatter $\tau_{\rm in}$. We rescale the total uncertainty in velocity dispersion to approximate the change in sample size; for example, using $0.1\tau$ mimics the statistical gain expected for a sample 100 times larger than the current one, where $\tau^2=\tau_{\rm obs}^2+\tau_{\rm in}^2$. Note that, for the noise-free mock data, the single-lens likelihood is constructed directly using the input lens mass distribution adopted to generate the mock velocity dispersion, rather than by marginalizing over $P(\alpha,b)$. 

\begin{figure}[!ht]
    \centering
    \includegraphics[width=\linewidth]{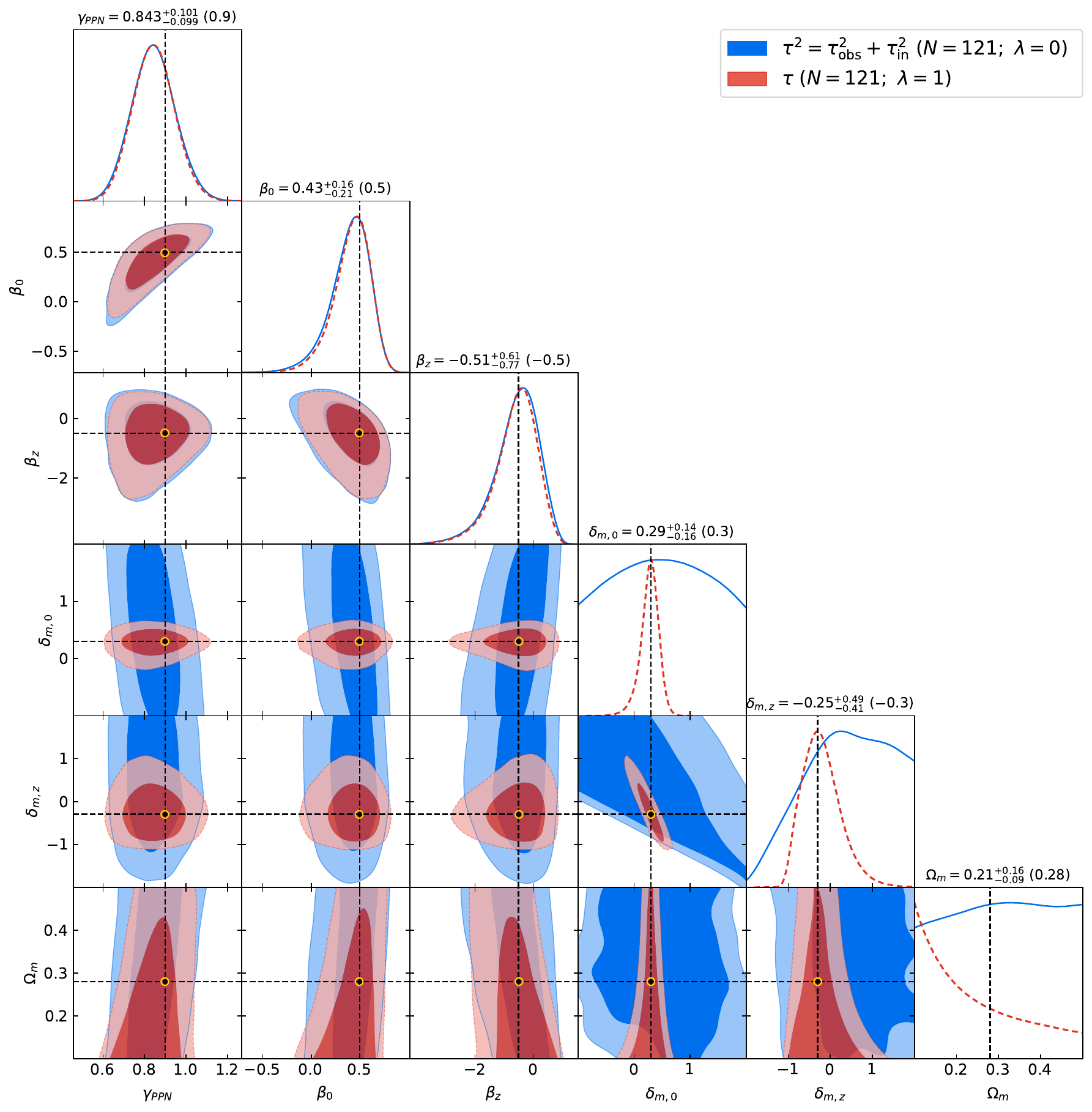}
    \caption{Recovery test with noiseless mock observed velocity dispersions. The mock data are generated with input values $\gamma_{\rm PPN}=0.9$, $\beta(z)=0.5-0.5z$, and $\delta_m(z)=0.3-0.3z$, as indicated by the black dashed lines and by the values in parentheses above the diagonal panels. The blue curves and contours show the inference with $\lambda=0$, while the red ones show the inference with $\lambda=1$, where the prior mean of the external convergence is specified by $\delta_m(z)=0.3-0.3z$. Yellow circles mark the best-fitting values, which almost coincide with the input values. Diagonal panels show the one-dimensional marginalized posteriors, with the median and 68\% credible interval for the red distributions reported above each panel. Off-diagonal panels show the two-dimensional marginalized posteriors, with contours enclosing the 68\% and 95\% credible regions. The gravity and anisotropy parameters are recovered consistently, whereas a sample of $\sim 100$ lenses provides only weak constraints on $\Omega_{\rm m}$ and external-convergence population parameters $\delta_{m,0}$ and $\delta_{m,z}$.}
    \label{figA5:mcmc_won}
\end{figure}

Fig.~\ref{figA5:mcmc_won} shows the baseline noise-free test for 121 lenses. In this test, $\tau_\beta$ is not included as a free parameter, because the input mock data contain no intrinsic anisotropy scatter. Additional runs in which $\tau_\beta$ is allowed to vary recover $\tau_\beta$ values consistent with zero. Fig.~\ref{figA5:mcmc_won} shows that the gravity and anisotropy parameters are recovered consistently. By contrast, the cosmological and external-convergence population parameters are weakly constrained for a sample of $\sim 100$ lenses. 

\restoregeometry

\subsection{Lens systems associated with redMaPPer clusters}\label{secA4:redmapper}

\begin{table}[h]
\caption{Thirteen lenses associated with redMaPPer clusters.}
\label{tabA2:redmapper}
\centering
\renewcommand{\arraystretch}{1.5}
\begin{tabular}{lccc}
\hline
Lens name & redMaPPer cluster & $\kappa_{\rm env}$ & $R_{\rm len}/r_{200}$ \\
\hline
J0157$-$0056 & RMJ015805.0$-$005853.0 & 0.0131 & 0.806 \\
J0159$-$0006 & RMJ015930.8$-$000616.8 & 0.2633 & 0.028 \\
J0216$-$0813$^{\rm c}$ & RMJ021652.5$-$081345.4 & --     & 0.000 \\
J0753$+$3416 & RMJ075325.2$+$341632.6 & 0.0168 & 0.487 \\
J0903$+$4116$^{\rm c}$ & RMJ090315.2$+$411609.1 & --     & 0.000 \\
J0920$+$3028 & RMJ092052.5$+$302740.3 & 0.0748 & 0.143 \\
J0935$-$0003$^{\rm c}$ & RMJ093543.9$-$000334.8 & --     & 0.000 \\
J0955$+$3014$^{\rm c}$ & RMJ095557.5$+$301450.9 & --     & 0.000 \\
J1032$+$5322 & RMJ103209.4$+$531912.9 & 0.0063 & 0.645 \\
J1143$-$0144$^{\rm c}$ & RMJ114329.6$-$014430.0 & --     & 0.000 \\
J1204$+$0358 & RMJ120435.1$+$035723.1 & 0.0593 & 0.241 \\
J1403$+$0006 & RMJ140320.0$+$000535.5 & 0.0266 & 0.341 \\
J2341$+$0000 & RMJ234059.8$+$000453.7 & 0.0067 & 0.888 \\
\hline
\end{tabular}
\footnotetext{\textit{Notes.} The table lists the lens name, the matched redMaPPer v6.3 cluster \cite{2015MNRAS.453...38R,2016ApJS..224....1R}, the environmental convergence $\kappa_{\rm env}$ at the lens position, and the projected lens--cluster separation normalized by $r_{200}$. A dash in the third column, also marked by the superscript $c$ in the first column, indicates lenses identified as central galaxies of the matched clusters, for which $\kappa_{\rm env}$ is set to zero to avoid double-counting the lens mass. The strong lenses SLACSJ2300$+$0022, BELLSJ1318$-$0104 and BELLSJ1611$+$1705 are projected near redMaPPer clusters but have large lens--cluster redshift offsets; these foreground or background clusters are therefore excluded from this table and treated as uncorrelated line-of-sight structures.}
\end{table}

\newgeometry{margin=0.5in,bottom=1.in}
\subsection{Lens sample properties}\label{secA5:lens_sample}

\setlength{\LTcapwidth}{\textwidth}
\renewcommand{\arraystretch}{1.2}
\small
\begin{longtable}{llccclcccc}
\caption{125 grade-A single lenses from the SLACS, S4TM and BELLS surveys \cite{2008ApJ...682..964B,2017ApJ...851...48S,2012ApJ...744...41B}.}
\label{tabA3:lens_sample}\\
\hline
Lens name & Survey & RA/Dec. & $z_{\rm d}$ & $z_{\rm s}$ & \mc{$\sigma_{\parallel,\rm obs}$} & Seeing & $s_{\rm fib}$ & $b_\sigma$ & $\langle\kappa_{\rm LSS}^2\rangle^{1/2}$ \\
\mc{(SDSS)}    &        & (J2000)   &      &           &      \mc{(km s$^{-1}$)}     & (arcsec) & (arcsec) &   &   \\
\mc{(1)}    &   \mc{(2)}     & (3)   &   (4)   &      (5)     &      \mc{(6)}     & (7) & (8) &  (9)  &  (10) \\
\hline
\endfirsthead

\multicolumn{10}{c}{\tablename~\thetable\ -- continued}\\
\hline
Lens name & Survey & RA/Dec. & $z_{\rm d}$ & $z_{\rm s}$ & \mc{$\sigma_{\parallel,\rm obs}$} & Seeing & $s_{\rm fib}$ & $b_\sigma$ & $\langle\kappa_{\rm LSS}^2\rangle^{1/2}$ \\
\mc{(SDSS)}    &        & (J2000)   &      &           &      \mc{(km s$^{-1}$)}     & (arcsec) & (arcsec) &   &   \\
\mc{(1)}    &   \mc{(2)}     & (3)   &   (4)   &      (5)     &      \mc{(6)}     & (7) & (8) &  (9)  &  (10) \\
\hline
\endhead

\hline
\multicolumn{10}{c}{\tablename~\thetable\ -- continued}\\
\endfoot
\hline
\endlastfoot
J0008$-$0004 & SLACS & 000802.96$-$000408.2 & 0.440 & 1.192 & 231.9$\pm$39.1 & 1.950 & 1.197 & 1.028 & 0.023 \\ 
J0029$-$0055 & SLACS & 002907.77$-$005550.5 & 0.227 & 0.931 & 201.7$\pm$18.2 & 1.840 & 1.175 & 1.028 & 0.013 \\ 
J0037$-$0942 & SLACS & 003753.22$-$094220.1 & 0.195 & 0.632 & 274.1$\pm$10.3 & 1.718 & 1.154 & 1.038 & 0.009 \\ 
J0044$+$0113 & SLACS & 004402.90$+$011312.5 & 0.120 & 0.197 & 254.3$\pm$12.2 & 1.692 & 1.151 & 1.033 & 0.003 \\ 
J0109$+$1500 & SLACS & 010933.74$+$150032.5 & 0.294 & 0.525 & 242.9$\pm$18.4 & 1.688 & 1.150 & 1.034 & 0.010 \\ 
J0157$-$0056 & SLACS & 015758.94$-$005626.1 & 0.514 & 0.924 & 228.3$\pm$30.6$^{*}$ & 1.745 & 0.919 & 1.041 & 0.020 \\ 
J0216$-$0813 & SLACS & 021652.53$-$081345.4 & 0.332 & 0.524 & 332.0$\pm$24.0 & 1.634 & 1.143 & 1.030 & 0.010 \\ 
J0252$+$0039 & SLACS & 025245.21$+$003958.4 & 0.280 & 0.982 & 172.1$\pm$10.7 & 2.072 & 1.225 & 1.021 & 0.015 \\ 
J0330$-$0020 & SLACS & 033012.14$-$002051.9 & 0.351 & 1.071 & 259.5$\pm$23.6 & 1.597 & 1.139 & 1.036 & 0.019 \\ 
J0405$-$0455 & SLACS & 040535.42$-$045552.4 & 0.075 & 0.810 & 168.8$\pm$7.3 & -- & 1.153 & 1.042 & 0.005 \\ 
J0728$+$3835 & SLACS & 072804.95$+$383525.7 & 0.206 & 0.688 & 219.6$\pm$10.9 & 1.230 & 1.140 & 1.037 & 0.010 \\ 
J0737$+$3216 & SLACS & 073728.44$+$321618.7 & 0.322 & 0.581 & 300.8$\pm$15.7 & 2.328 & 1.292 & 1.028 & 0.011 \\ 
J0822$+$2652 & SLACS & 082242.32$+$265243.5 & 0.241 & 0.594 & 254.7$\pm$15.3 & 1.935 & 1.194 & 1.036 & 0.010 \\ 
J0841$+$3824 & SLACS & 084128.81$+$382413.7 & 0.116 & 0.657 & 219.8$\pm$8.1 & 1.470 & 1.130 & 1.054 & 0.006 \\ 
J0903$+$4116 & SLACS & 090315.19$+$411609.1 & 0.430 & 1.064 & 203.1$\pm$24.1 & 1.508 & 1.132 & 1.024 & 0.021 \\ 
J0912$+$0029 & SLACS & 091205.31$+$002901.2 & 0.164 & 0.324 & 309.7$\pm$11.8 & 2.802 & 1.438 & 1.044 & 0.005 \\ 
J0935$-$0003 & SLACS & 093543.93$-$000334.8 & 0.347 & 0.467 & 344.8$\pm$41.0 & 2.467 & 1.332 & 1.023 & 0.009 \\ 
J0936$+$0913 & SLACS & 093600.77$+$091335.8 & 0.190 & 0.588 & 238.4$\pm$11.5 & 1.438 & 1.129 & 1.031 & 0.009 \\ 
J0946$+$1006 & SLACS & 094656.68$+$100652.8 & 0.222 & 0.609 & 256.6$\pm$20.5 & 1.237 & 1.139 & 1.017 & 0.010 \\ 
J0955$+$0101 & SLACS & 095519.72$+$010144.4 & 0.111 & 0.316 & 197.0$\pm$12.6 & 1.612 & 1.141 & 1.090 & 0.004 \\ 
J0956$+$5100 & SLACS & 095629.78$+$510006.4 & 0.241 & 0.470 & 297.4$\pm$16.2 & 1.653 & 1.146 & 1.038 & 0.008 \\ 
J0959$+$0410 & SLACS & 095944.07$+$041017.0 & 0.126 & 0.535 & 202.8$\pm$12.3 & 1.822 & 1.172 & 1.049 & 0.006 \\ 
J0959$+$4416 & SLACS & 095900.96$+$441639.4 & 0.237 & 0.532 & 238.5$\pm$19.8 & 1.593 & 1.139 & 1.024 & 0.009 \\ 
J1016$+$3859 & SLACS & 101622.86$+$385903.3 & 0.168 & 0.439 & 250.6$\pm$13.5 & 2.237 & 1.267 & 1.027 & 0.006 \\ 
J1020$+$1122 & SLACS & 102026.54$+$112241.1 & 0.282 & 0.553 & 274.0$\pm$18.6 & 1.285 & 1.134 & 1.031 & 0.010 \\ 
J1023$+$4230 & SLACS & 102332.26$+$423001.8 & 0.191 & 0.696 & 254.8$\pm$13.7 & 1.850 & 1.177 & 1.027 & 0.010 \\ 
J1029$+$0420 & SLACS & 102922.94$+$042001.8 & 0.104 & 0.615 & 208.0$\pm$8.6 & 2.140 & 1.242 & 1.064 & 0.006 \\ 
J1032$+$5322 & SLACS & 103235.84$+$532234.9 & 0.133 & 0.329 & 293.0$\pm$15.0 & 2.030 & 1.215 & 1.078 & 0.004 \\ 
J1100$+$5329 & SLACS & 110024.39$+$532913.7 & 0.317 & 0.858 & 189.7$\pm$25.0 & 1.940 & 1.195 & 1.053 & 0.015 \\ 
J1103$+$5322 & SLACS & 110308.21$+$532228.2 & 0.158 & 0.735 & 199.2$\pm$11.9 & 1.940 & 1.195 & 1.074 & 0.009 \\ 
J1106$+$5228 & SLACS & 110646.16$+$522837.7 & 0.096 & 0.407 & 263.8$\pm$10.2 & 1.940 & 1.195 & 1.048 & 0.004 \\ 
J1112$+$0826 & SLACS & 111250.60$+$082610.4 & 0.273 & 0.630 & 298.9$\pm$21.4 & 3.020 & 1.511 & 1.035 & 0.011 \\ 
J1134$+$6027 & SLACS & 113405.88$+$602713.2 & 0.153 & 0.474 & 232.8$\pm$10.2 & 2.205 & 1.258 & 1.030 & 0.006 \\ 
J1142$+$1001 & SLACS & 114257.35$+$100111.8 & 0.222 & 0.504 & 212.9$\pm$22.1 & 1.927 & 1.192 & 1.022 & 0.008 \\ 
J1143$-$0144 & SLACS & 114329.64$-$014430.0 & 0.106 & 0.402 & 261.7$\pm$5.0 & 1.876 & 1.182 & 1.031 & 0.004 \\ 
J1153$+$4612 & SLACS & 115310.79$+$461205.3 & 0.180 & 0.875 & 210.9$\pm$14.7 & 1.607 & 1.140 & 1.024 & 0.010 \\ 
J1204$+$0358 & SLACS & 120444.07$+$035806.4 & 0.164 & 0.631 & 246.2$\pm$15.7 & 1.410 & 1.129 & 1.017 & 0.008 \\ 
J1205$+$4910 & SLACS & 120540.44$+$491029.4 & 0.215 & 0.481 & 295.9$\pm$13.8 & 2.268 & 1.275 & 1.040 & 0.008 \\ 
J1213$+$6708 & SLACS & 121340.58$+$670829.0 & 0.123 & 0.640 & 278.4$\pm$10.0 & 2.293 & 1.282 & 1.033 & 0.007 \\ 
J1218$+$0830 & SLACS & 121826.70$+$083050.3 & 0.135 & 0.717 & 222.2$\pm$9.5 & 1.960 & 1.199 & 1.038 & 0.008 \\ 
J1250$+$0523 & SLACS & 125028.26$+$052349.1 & 0.232 & 0.795 & 246.3$\pm$14.9 & 2.103 & 1.232 & 1.017 & 0.012 \\ 
J1251$-$0208 & SLACS & 125135.71$-$020805.2 & 0.224 & 0.784 & 211.9$\pm$23.0 & 2.290 & 1.281 & 1.064 & 0.012 \\ 
J1402$+$6321 & SLACS & 140228.22$+$632133.3 & 0.205 & 0.481 & 267.6$\pm$15.3 & 2.370 & 1.304 & 1.034 & 0.008 \\ 
J1403$+$0006 & SLACS & 140329.49$+$000641.3 & 0.189 & 0.473 & 205.5$\pm$15.5 & 1.722 & 1.155 & 1.031 & 0.007 \\ 
J1416$+$5136 & SLACS & 141622.34$+$513630.4 & 0.299 & 0.811 & 230.4$\pm$22.4 & 2.183 & 1.253 & 1.033 & 0.014 \\ 
J1420$+$6019 & SLACS & 142015.85$+$601914.8 & 0.063 & 0.535 & 198.9$\pm$4.0 & 2.083 & 1.228 & 1.058 & 0.003 \\ 
J1430$+$4105 & SLACS & 143004.10$+$410557.1 & 0.285 & 0.575 & 294.3$\pm$29.7 & 1.583 & 1.138 & 1.033 & 0.011 \\ 
J1432$+$6317 & SLACS & 143213.34$+$631703.8 & 0.123 & 0.664 & 193.2$\pm$8.2 & 1.323 & 1.131 & 1.018 & 0.007 \\ 
J1436$-$0000 & SLACS & 143627.54$-$000029.1 & 0.285 & 0.805 & 217.6$\pm$16.9 & 1.925 & 1.192 & 1.036 & 0.014 \\ 
J1443$+$0304 & SLACS & 144319.62$+$030408.2 & 0.134 & 0.419 & 221.9$\pm$10.9 & 2.908 & 1.473 & 1.049 & 0.005 \\ 
J1451$-$0239 & SLACS & 145128.19$-$023936.4 & 0.125 & 0.520 & 212.7$\pm$13.2 & 2.360 & 1.301 & 1.016 & 0.006 \\ 
J1525$+$3327 & SLACS & 152506.70$+$332747.4 & 0.358 & 0.717 & 302.6$\pm$29.2 & 1.447 & 1.129 & 1.051 & 0.014 \\ 
J1531$-$0105 & SLACS & 153150.08$-$010545.6 & 0.160 & 0.744 & 259.2$\pm$9.3 & 1.813 & 1.170 & 1.042 & 0.009 \\ 
J1538$+$5817 & SLACS & 153812.94$+$581709.7 & 0.143 & 0.531 & 191.4$\pm$11.8 & 2.707 & 1.407 & 1.029 & 0.007 \\ 
J1621$+$3931 & SLACS & 162132.99$+$393144.6 & 0.245 & 0.602 & 228.2$\pm$20.3 & 1.807 & 1.169 & 1.038 & 0.010 \\ 
J1627$-$0053 & SLACS & 162746.45$-$005357.6 & 0.208 & 0.524 & 286.1$\pm$13.1 & 1.850 & 1.177 & 1.026 & 0.008 \\ 
J1630$+$4520 & SLACS & 163028.16$+$452036.3 & 0.248 & 0.793 & 252.2$\pm$16.1 & 1.543 & 1.134 & 1.027 & 0.013 \\ 
J1636$+$4707 & SLACS & 163602.62$+$470729.6 & 0.228 & 0.674 & 222.9$\pm$15.2 & 1.353 & 1.130 & 1.034 & 0.011 \\ 
J2238$-$0754 & SLACS & 223840.20$-$075456.0 & 0.137 & 0.713 & 222.3$\pm$10.7 & 1.548 & 1.135 & 1.037 & 0.008 \\ 
J2300$+$0022 & SLACS & 230053.14$+$002237.9 & 0.228 & 0.463 & 267.2$\pm$18.1 & 1.933 & 1.193 & 1.032 & 0.008 \\ 
J2303$+$1422 & SLACS & 230321.72$+$142217.9 & 0.155 & 0.517 & 248.7$\pm$15.3 & 1.483 & 1.131 & 1.046 & 0.007 \\ 
J2321$-$0939 & SLACS & 232120.93$-$093910.3 & 0.082 & 0.532 & 238.8$\pm$8.0 & 1.580 & 1.137 & 1.033 & 0.004 \\ 
J2341$+$0000 & SLACS & 234111.57$+$000018.7 & 0.186 & 0.807 & 223.8$\pm$13.7 & 1.417 & 1.129 & 1.055 & 0.010 \\ 
J0143$-$1006 & S4TM & 014356.58$-$100633.7 & 0.221 & 1.105 & 207.8$\pm$18.7 & 2.053 & 1.220 & 1.032 & 0.014 \\ 
J0159$-$0006 & S4TM & 015930.14$-$000612.4 & 0.158 & 0.748 & 212.8$\pm$18.1 & 1.850 & 1.177 & 1.022 & 0.009 \\ 
J0324$+$0045 & S4TM & 032415.50$+$004505.5 & 0.321 & 0.920 & 181.5$\pm$20.5 & 1.757 & 1.160 & 1.027 & 0.016 \\ 
J0324$-$0110 & S4TM & 032454.50$-$011029.1 & 0.446 & 0.624 & 307.0$\pm$38.2 & 1.234 & 1.139 & 1.037 & 0.013 \\ 
J0753$+$3416 & S4TM & 075346.21$+$341633.6 & 0.137 & 0.963 & 208.7$\pm$12.6 & 1.367 & 1.129 & 1.026 & 0.009 \\ 
J0754$+$1927 & S4TM & 075428.52$+$192728.1 & 0.153 & 0.740 & 192.2$\pm$16.5 & 1.907 & 1.188 & 1.019 & 0.008 \\ 
J0757$+$1956 & S4TM & 075748.99$+$195616.3 & 0.121 & 0.833 & 206.2$\pm$11.7 & 1.087 & 1.170 & 1.022 & 0.007 \\ 
J0826$+$5630 & S4TM & 082639.86$+$563036.0 & 0.132 & 1.291 & 163.9$\pm$8.6 & 1.797 & 1.167 & 1.025 & 0.010 \\ 
J0847$+$2348 & S4TM & 084727.69$+$234819.5 & 0.155 & 0.533 & 200.7$\pm$15.8 & 1.120 & 1.161 & 1.019 & 0.007 \\ 
J0851$+$0505 & S4TM & 085141.89$+$050507.0 & 0.128 & 0.637 & 176.3$\pm$11.9 & 1.337 & 1.130 & 1.023 & 0.007 \\ 
J0920$+$3028 & S4TM & 092048.28$+$302818.4 & 0.288 & 0.392 & 302.6$\pm$17.2 & 1.363 & 1.129 & 1.021 & 0.007 \\ 
J0955$+$3014 & S4TM & 095557.49$+$301450.9 & 0.321 & 0.467 & 259.7$\pm$28.6 & 1.217 & 1.142 & 1.038 & 0.009 \\ 
J0956$+$5539 & S4TM & 095654.84$+$553947.3 & 0.196 & 0.848 & 188.9$\pm$11.6 & 2.112 & 1.235 & 1.017 & 0.011 \\ 
J1010$+$3124 & S4TM & 101026.80$+$312417.6 & 0.167 & 0.425 & 220.8$\pm$12.1 & 1.225 & 1.141 & 1.036 & 0.006 \\ 
J1041$+$0112 & S4TM & 104122.85$+$011224.2 & 0.101 & 0.217 & 198.1$\pm$7.6 & 1.662 & 1.147 & 1.026 & 0.003 \\ 
J1048$+$1313 & S4TM & 104809.40$+$131352.9 & 0.133 & 0.668 & 198.3$\pm$10.1 & 0.956 & 1.223 & 1.050 & 0.007 \\ 
J1051$+$4439 & S4TM & 105109.41$+$443908.5 & 0.163 & 0.538 & 216.1$\pm$16.2 & 1.210 & 1.143 & 1.033 & 0.007 \\ 
J1056$+$4141 & S4TM & 105657.61$+$414114.6 & 0.134 & 0.832 & 157.2$\pm$10.3 & 1.370 & 1.129 & 1.026 & 0.008 \\ 
J1101$+$1523 & S4TM & 110113.13$+$152339.6 & 0.178 & 0.517 & 274.5$\pm$15.0 & 1.397 & 1.129 & 1.040 & 0.008 \\ 
J1116$+$0729 & S4TM & 111641.66$+$072945.6 & 0.170 & 0.686 & 190.3$\pm$11.8 & 1.247 & 1.138 & 1.030 & 0.009 \\ 
J1127$+$2312 & S4TM & 112738.70$+$231244.4 & 0.130 & 0.361 & 231.9$\pm$9.7 & 1.810 & 1.169 & 1.023 & 0.005 \\ 
J1137$+$1818 & S4TM & 113728.61$+$181812.4 & 0.124 & 0.463 & 222.3$\pm$9.0 & 1.110 & 1.164 & 1.023 & 0.005 \\ 
J1142$+$2509 & S4TM & 114238.23$+$250905.5 & 0.164 & 0.659 & 158.9$\pm$10.1 & 1.140 & 1.156 & 1.024 & 0.008 \\ 
J1144$+$0436 & S4TM & 114440.13$+$043650.5 & 0.104 & 0.255 & 201.0$\pm$14.9 & 1.437 & 1.129 & 1.028 & 0.003 \\ 
J1213$+$2930 & S4TM & 121303.72$+$293022.4 & 0.091 & 0.595 & 233.5$\pm$7.8 & 1.337 & 1.130 & 1.044 & 0.005 \\ 
J1301$+$0834 & S4TM & 130126.88$+$083425.2 & 0.090 & 0.533 & 180.4$\pm$8.2 & 1.513 & 1.132 & 1.058 & 0.005 \\ 
J1330$+$1750 & S4TM & 133031.40$+$175040.5 & 0.207 & 0.372 & 249.5$\pm$12.3 & 1.293 & 1.133 & 1.036 & 0.006 \\ 
J1403$+$3309 & S4TM & 140309.67$+$330917.8 & 0.062 & 0.772 & 190.8$\pm$6.5 & 1.063 & 1.178 & 1.030 & 0.004 \\ 
J1430$+$6104 & S4TM & 143034.77$+$610404.8 & 0.169 & 0.654 & 182.5$\pm$15.3 & 1.796 & 1.167 & 1.031 & 0.009 \\ 
J1433$+$2835 & S4TM & 143351.63$+$283516.4 & 0.091 & 0.411 & 231.2$\pm$6.6 & 1.110 & 1.164 & 1.018 & 0.004 \\ 
J1541$+$3642 & S4TM & 154122.27$+$364231.7 & 0.141 & 0.739 & 193.7$\pm$11.5 & 1.743 & 1.158 & 1.019 & 0.008 \\ 
J1543$+$2202 & S4TM & 154339.94$+$220223.3 & 0.268 & 0.397 & 286.7$\pm$16.9 & 1.220 & 1.141 & 1.031 & 0.007 \\ 
J1550$+$2020 & S4TM & 155010.62$+$202013.5 & 0.135 & 0.350 & 244.5$\pm$9.7 & 1.517 & 1.132 & 1.042 & 0.005 \\ 
J1553$+$3004 & S4TM & 155316.14$+$300425.7 & 0.160 & 0.566 & 196.4$\pm$15.5 & 1.037 & 1.187 & 1.021 & 0.007 \\ 
J1607$+$2147 & S4TM & 160740.50$+$214711.0 & 0.209 & 0.487 & 193.7$\pm$16.3 & 1.386 & 1.129 & 1.022 & 0.008 \\ 
J1633$+$1441 & S4TM & 163344.16$+$144154.9 & 0.128 & 0.580 & 234.4$\pm$9.1 & 1.122 & 1.160 & 1.028 & 0.006 \\ 
J2309$-$0039 & S4TM & 230946.36$-$003912.9 & 0.291 & 1.005 & 185.1$\pm$14.0 & 1.690 & 1.150 & 1.017 & 0.016 \\ 
J2324$+$0105 & S4TM & 232427.77$+$010558.5 & 0.190 & 0.278 & 245.1$\pm$15.1 & 1.723 & 1.155 & 1.061 & 0.005 \\ 
J0151$+$0049$^{a}$ & BELLS & 015107.37$+$004909.0 & 0.517 & 1.364 & 0.0$\pm$0.0 & -- & 0.879 & 1.053 & 0.027 \\ 
J0747$+$4448 & BELLS & 074734.75$+$444859.3 & 0.436 & 0.897 & 232.1$\pm$39.6 & -- & 0.879 & 1.051 & 0.019 \\ 
J0747$+$5055 & BELLS & 074724.12$+$505537.5 & 0.438 & 0.898 & 240.2$\pm$39.6 & -- & 0.879 & 1.036 & 0.019 \\ 
J0801$+$4727 & BELLS & 080105.30$+$472749.6 & 0.483 & 1.518 & 117.0$\pm$24.8 & -- & 0.879 & 1.016 & 0.027 \\ 
J0830$+$5116 & BELLS & 083049.73$+$511631.8 & 0.530 & 1.332 & 225.9$\pm$33.3 & -- & 0.879 & 1.038 & 0.026 \\ 
J0944$-$0147 & BELLS & 094427.47$-$014742.4 & 0.539 & 1.179 & 171.0$\pm$43.0 & -- & 0.879 & 1.026 & 0.025 \\ 
J1159$-$0007$^{a}$ & BELLS & 115944.63$-$000728.2 & 0.332 & 1.346 & 850.0$\pm$-3.0 & 1.608 & 0.877 & 1.018 & 0.028 \\ 
J1215$+$0047$^{a}$ & BELLS & 121504.44$+$004726.0 & 0.642 & 1.297 & 748.0$\pm$64.0 & 1.880 & 0.963 & 1.042 & 0.028 \\ 
J1221$+$3806 & BELLS & 122151.92$+$380610.5 & 0.535 & 1.284 & 219.0$\pm$48.8 & 1.167 & 0.773 & 1.028 & 0.026 \\ 
J1234$-$0241 & BELLS & 123427.99$-$024129.6 & 0.490 & 1.016 & 128.1$\pm$35.3 & -- & 0.879 & 1.037 & 0.021 \\ 
J1318$-$0104 & BELLS & 131829.39$-$010421.6 & 0.659 & 1.396 & 204.4$\pm$29.5 & 1.260 & 0.790 & 1.030 & 0.030 \\ 
J1337$+$3620 & BELLS & 133751.31$+$362018.1 & 0.564 & 1.182 & 245.3$\pm$39.1 & 1.875 & 0.961 & 1.024 & 0.025 \\ 
J1349$+$3612 & BELLS & 134910.30$+$361239.7 & 0.440 & 0.893 & 198.3$\pm$22.7 & -- & 0.879 & 1.031 & 0.018 \\ 
J1352$+$3216$^{a}$ & BELLS & 135218.99$+$321651.8 & 0.463 & 1.034 & 162.7$\pm$24.6 & 1.515 & 0.851 & 1.019 & 0.021 \\ 
J1541$+$1812 & BELLS & 154118.56$+$181235.1 & 0.560 & 1.113 & 187.4$\pm$27.2 & 1.208 & 0.780 & 1.035 & 0.024 \\ 
J1542$+$1629 & BELLS & 154246.33$+$162951.8 & 0.352 & 1.023 & 217.5$\pm$17.5 & 1.126 & 0.767 & 1.034 & 0.018 \\ 
J1545$+$2748 & BELLS & 154503.57$+$274805.3 & 0.522 & 1.289 & 237.1$\pm$36.2 & 1.794 & 0.935 & 1.046 & 0.026 \\ 
J1601$+$2138 & BELLS & 160113.27$+$213833.9 & 0.543 & 1.446 & 278.4$\pm$32.3 & 1.920 & 0.976 & 1.022 & 0.028 \\ 
J1611$+$1705 & BELLS & 161109.80$+$170526.6 & 0.477 & 1.211 & 99.6$\pm$22.4 & 1.660 & 0.893 & 1.021 & 0.024 \\ 
J1631$+$1854 & BELLS & 163150.33$+$185404.1 & 0.408 & 1.086 & 293.7$\pm$16.7 & 1.763 & 0.925 & 1.020 & 0.020 \\ 
J1637$+$1439 & BELLS & 163714.58$+$143930.1 & 0.391 & 0.874 & 214.9$\pm$35.3 & 1.887 & 0.965 & 1.053 & 0.017 \\ 
J2122$+$0409 & BELLS & 212252.04$+$040935.5 & 0.626 & 1.452 & 308.1$\pm$48.9 & 1.832 & 0.947 & 1.028 & 0.030 \\ 
J2125$+$0411 & BELLS & 212510.67$+$041131.6 & 0.363 & 0.978 & 253.9$\pm$20.6 & 1.316 & 0.801 & 1.037 & 0.018 \\ 
J2303$+$0037 & BELLS & 230335.17$+$003703.2 & 0.458 & 0.936 & 260.1$\pm$30.0 & 1.856 & 0.955 & 1.032 & 0.020 \\ 


\end{longtable}
\noindent\textit{Notes.} 
Column (1) shows the lens name. The superscript $a$ marks the four outliers excluded when defining the final sample; their observed velocity dispersions differ from the model predictions by more than $100~{\rm km~s^{-1}}$ in the initial lensing--dynamical consistency check.
Column (2) gives the survey programme from which the lens is selected. 
Column (3) gives the sky coordinates of the lens in J2000. 
Columns (4) and (5) are the lens and source redshifts, respectively. 
Column (6) gives the observed stellar velocity dispersion reported in the SDSS DR17 main spectroscopic catalogue. In this column, the superscript $*$ indicates that the velocity dispersion for J0157$-$0056 is taken from the BOSS spectroscopic measurement.
Column (7) shows the FWHM of the median seeing during spectroscopic observations. A dash in this column indicates that the seeing information is unavailable; for these systems, the mean seeing of the lenses from the corresponding survey is adopted.
Column (8) provides the effective Gaussian width of the aperture weighting function, which accounts for the combined effects of seeing and fibre size and is used to compute the single-aperture velocity dispersion.
Column (9) shows the projection bias $b_\sigma=1.015q^{-0.07}_\star$ in the dynamical modelling based on the broken power-law lens mass model, where $q_\star$ is the axis ratio of the projected lens light distribution. 
Column (10) gives the model-predicted scatter in external convergence from uncorrelated line-of-sight large-scale structure.
\restoregeometry

\end{appendices}


\end{document}